\documentclass[12pt,english,article,amsmath,amssymb]{article}
\usepackage{geometry}
\usepackage{graphicx,graphics}
\usepackage{dcolumn}
\usepackage{amsmath,amssymb,amsfonts}
\usepackage{latexsym,verbatim}
\usepackage{bm,bbm}
\usepackage{color}
\usepackage{ulem}

\usepackage[T1]{fontenc}
\usepackage{dcolumn}
\usepackage{color}
\usepackage{babel}
\usepackage{braket}

\usepackage[percent]{overpic}

\newcommand{\bq}{{\bm q}}
\newcommand{\br}{{\bm r}}

\newcommand{\kF}{k_{\mathrm{F}}}
\newcommand{\vD}{v_{\mathrm{D}}}

\newcommand{\Real}{\mathrm{Re}}
\newcommand{\Imag}{\mathrm{Im}}
\newcommand{\openone}{\mathbbm{1}}
\definecolor{dgreen}{rgb}{0,0.5,0}
\begin{document}

\title{
Effect of dilute impurities on short graphene Josephson junctions
}
\author{Francesco M.D. Pellegrino$^{1,2,3,4}$\footnote{Corresponding author: francesco.pellegrino@dfa.unict.it}, 
Giuseppe Falci$^{1,2,3,4}$ and Elisabetta Paladino$^{1,2,3,4}$}
\date{}
\maketitle
\noindent$^1$ Dipartimento di Fisica e Astronomia "Ettore Majorana", Universit\`{a} di Catania,\\ Via S.~Sofia 64, I-95123 Catania, Italy. \\
$^2$ INFN, Sez. Catania, I-95123 Catania, Italy. \\
$^3$ CNR-IMM, Via S. Sofia 64, I-95123 Catania, Italy.\\
$^4$ CSFNSM, Via S. Sofia 64, I-95123 Catania, Italy.\\
\maketitle

\begin{abstract}
We study the effect of a dilute homogeneous spatial distribution of non-magnetic impurities on the equilibrium supercurrent sustained by a ballistic graphene Josephson junction in the short junction limit. Within the Dirac-Bogoliubov-de Gennes approach and modeling impurities by the Anderson model
we derive the supercurrent and its equilibrium power spectrum.
We find a modification of the current-phase relation with a reduction of the skewness induced by disorder, and a nonmonotonic temperature dependence of the critical current. The potentialities of the supercurrent power spectrum 
for accurate spectroscopy of the hybridized Andreev bound states-impurities spectrum are highlighted. In the low temperature limit, the supercurrent
zero frequency thermal noise directly probes the spectral function at the Fermi energy.  
\end{abstract}

The future of quantum technologies lies in hybrid systems achieving multitasking potentialities by combining different physical components with complementary functionalities~\cite{Laucht_2021_Nanotech,deLeon_Science_2021}. In particular, devices based on hybrid Josephson junctions have opened up new possibilities to engineer noise protected qubits being at the same time easily tunable via electrical ports~\cite{Gyenis_PRXQuantum_2021}. Gate tunable superconducting qubits, so-called gatemons, have been successfully implemented with semiconducting nanowires~\cite{larsen_prl_2015,delange_prl_2015}, InAs Josephson junctions~\cite{Nichele_PRL_2020,Kringhoj_PRB_2018,Kringhoj_PRL_2020}, 2D materials~\cite{Lee_NanoLett_2019}, van der Waals heterostructures~\cite{Abhinandan_NanoLett_2021}
and graphene~\cite{Casparis_NatNanotech_2018, wang_natnanotech_2019}. Their promising characteristics are reduced dissipative losses, crosstalk and compatibility with high magnetic fields~\cite{schmidt_natcomm_2018,kroll_natcomm_2018}. An exciting perspective is creating fault-tolerant topological qubits based on Majorana zero modes ~\cite{shabani_prb_2016,Nayak_RevModPhys_2008}.
A fundamental step towards these achievements has been the realization of high-quality graphene superconductor heterostructures with clean interfaces obtained by encapsulating graphene in hexagonal boron nitride (hBN) with one-dimensional edge contacts to superconducting leads~\cite{dean_natnano_2010,mayorov_nanolett_2011, wang_science_2013}. These heterostructures show ballistic transport of Cooper pairs over micron-scale lengths, gate-tunable supercurrents that persist at large parallel magnetic fields~
\cite{calado_natnanotech_2015,benshalom_natphys_2016,borzenets_prl_2016}, and different features of 2D Andreev physics \cite{allen_natphys_2016,amet_science_2016,bretheau_natphys_2017}.
In addition, the extremely low specific-heat of graphene embedded in hBN allowed the realization of high sensitive GJJ-based microwave bolometers enabling circuit quantum electrodynamics applications~\cite{lee-efetov_nat_2020,kokkoniemi_nat_2020}.
Single near-infrared photon detection has also been proven by coupling photons to localized surface plasmons of a GJJ  which can be readily integrated into future JJ-based computing architectures as a high-speed, low-power optical interconnects~\cite{walsh-efetov_science_2021}. 
 GJJ is an excellent platform to realize exotic quantum states, recently long-lived Floquet-Andreev states have been generated by applying continuous microwave light without significant heating~\cite{park_nature_2022}.

The unifying microscopic description of the Josephson effect in these heterostructures results from proximity effect and constructive interference between Andreev processes at the two N/S interfaces leading to coherent electron-hole superpositions, known as Andreev bound states (ABSs).
In the short junction regime, the current-phase relation (CPR) resulting from the phase-dependence of the ABSs spectrum and density of states, differs from the sinusoidal CPR of tunnel Josephson junctions \cite{golubov_rmp_2004,Bretheau_PRX_2013} showing a skewness intrinsically related to the microscopic characteristics of the junctions as the number of transmitting channels and their transparency, and dependent on gate voltage and temperature \cite{english_prb_2016,nanda_nanol_2017}. 
Recently, concomitant measurements of CPR and Andreev bound state spectrum  in a highly transmissive  InAs Josephson junction~\cite{Nichele_PRL_2020} highlighted the potentialities of hybrid planar JJ as  sensors of fundamental phenomena 
occurring in heterostructures. Tunneling spectroscopy measurements in GJJ revealed the possible presence of microscopic quantum dots weakly coupled to the proximitized graphene, that behave as energy filters in tunneling process~\cite{wang_natnanotech_2019,wang_prb_2018}.  
Whether these impurities may influence the supercurrent of GJJ has not been yet established. 
On the other side it has been predicted that
carrier density fluctuations of the graphene channel due carrier traps in the nearby substrate~\cite{pellegrino_jstat_2019,kumar_apl_2021}  may induce critical current fluctuations with $1/f$ spectrum~\cite{pellegrino_commphys_2020,pellegrino_epjst_2021,paladino_rmp_2014}.
An alternative mechanism, related to variation of the proximity induced gap in the graphene junction fabricated using hBN encapsulation, has been reported~\cite{haque_scirep_2021}.

In this work, motivated by these observations,  we investigate the effect of a dilute ensemble of non-magnetic localized impurities on the equilibrium supercurrent in a ballistic GJJ, employing an analytical approach based on the Dirac-Bogoliubov-de Gennes model~\cite{titov_prb_2006}. 
In particular, we focus on the short channel limit~\cite{golubov_rmp_2004}, where the junction length is much smaller than the coherence length of the superconductors.
Impurities are modeled by the Anderson model~\cite{anderson_prev_1961}, which has been used to study the effect of adatoms on the graphene electron system~\cite{farjam_prb_2011,yuan_prb_2010,wehling_prl_2010,skrypnyk_jpcondmat_2013,barth_preprint_2021}.
Mu\~noz {\it et al.}~\cite{munoz_prb_2015} have recently investigated, using a self-consistent tight-binding approach, the influence of ripples~\cite{komatsu_prb_2012,choi_natcomm_2013} and localized defects, described as Lifshitz impurities~\cite{skrypnyk_prb_2011}, on an intermediate length GJJ, where multiple ABSs occur at zero temperature. 
A Lifshitz impurity modifies the on-site energy at its location in the corresponding tight-binding
Hamiltonian. In the dilute limit this type of disorder introduces a finite width to the Andreev peaks in the density of states, in agreement with the results obtained for a generic SNS junction with quasiclassical methods~\cite{bespalov_prb_2018}.
Contrary to the Lifshitz model, the Anderson model includes the  possibility of electron transfer from the host to some energy level that belongs to the adsorbed atom~\cite{skrypnyk_jpcondmat_2013}.

We derive the CPR of the disordered GJJ and demonstrate that dilute impurities are responsible for
a peculiar forward skewness effect accompanied by the reduction of the critical current. Both quantities display a characteristic nonmonotonic temperature dependence rooted in the hybridized ABS-impurities energies.   
These results are complemented by the derivation of supercurrent power spectrum 
which allows to perform spectroscopy of impurity levels with energies close to the Fermi energy. In the static limit and at very low temperatures, the supercurrent noise displays a linear temperature dependence, resembling thermal noise, with a slope related to energy distribution of the impurity states.
These results highlight the potentialities of short GJJ as highly sensitive detectors of microscopic defects spectral
characteristics via measurements of the supercurrent and its thermal equilibrium noise.

\begin{figure}[ht]
\centering
\centering
\includegraphics[width=0.79\columnwidth]{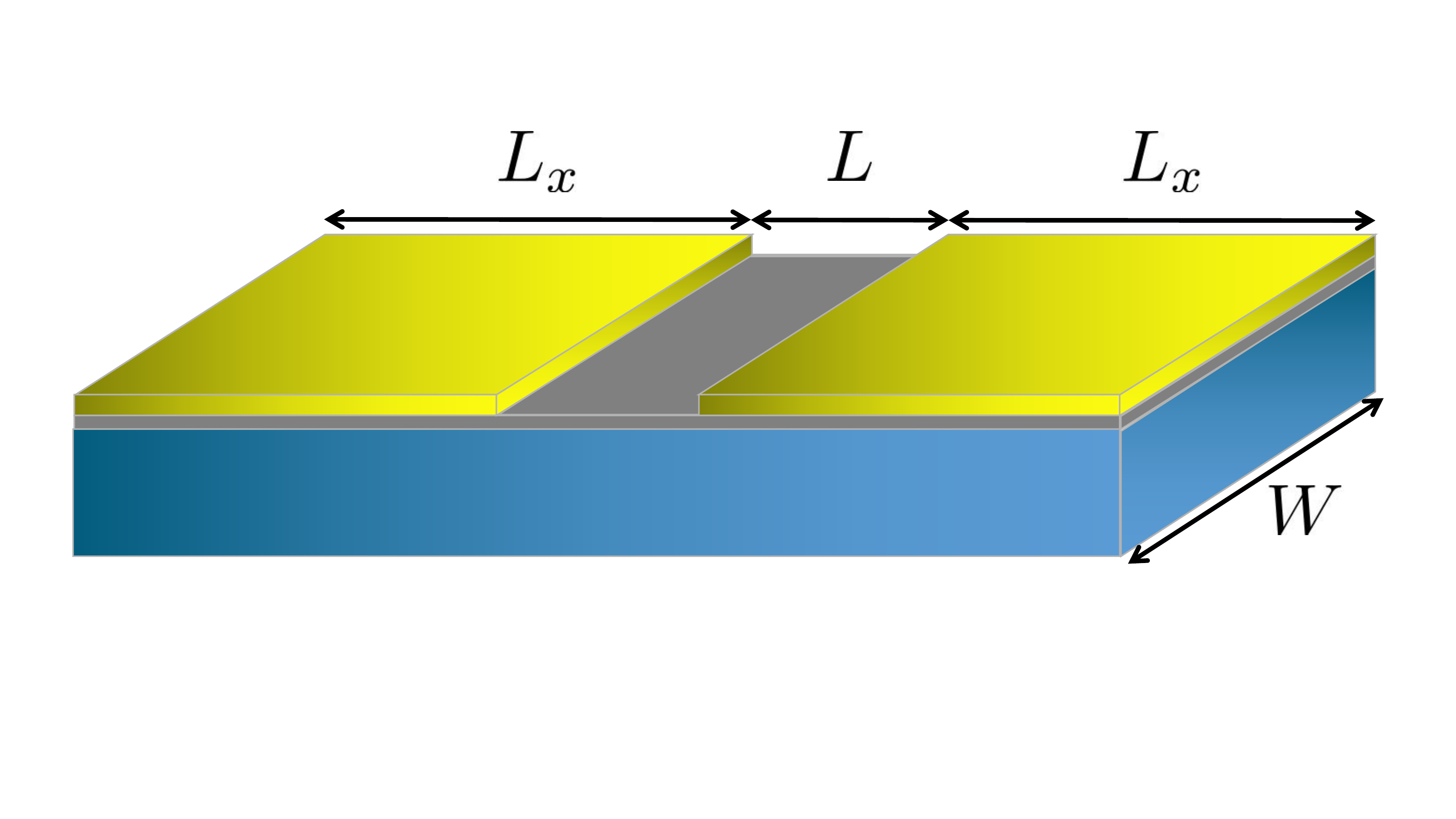}\\
\caption{Schematic of the device. From bottom to top there are a substrate (blue), a monolayer graphene  (gray) and two superconducting electrodes (yellow). 
The uncovered gray region represents the stripe in normal phase and yellow sides are the regions covered by superconductors.
Here, $L$ represents the junction channel, $L_x$ is the lateral size of each superconducting electrode, and $W$ is the length of the device along the invariant direction.
}
\label{fig:scheme}
\end{figure}

{\bf Results  and Discussion}\\
{\bf Model.} The system considered in this work, schematically shown in  Fig.~\ref{fig:scheme}, consists of a graphene layer (gray) partially covered by two superconducting electrodes (yellow), and deposited on a substrate (blue).
We model the GJJ in the ballistic regime within the  Dirac-Bogoliubov-de Gennes (D-BdG) approach, where  superconducting metal stripes induce  very large doping and superconductivity by proximity effect in the underlying graphene layer~ \cite{titov_prb_2006,takane_jpsj_2011,takane_jpsj_2012,alidoust_prr_2020,Hagymasi_PRB_2010}.
The D-BdG Hamiltonian reads
\begin{equation}\label{eq:HDBdG_CB}
\hat{\cal H}_{\rm D-BdG} = \sum_{\zeta=\pm} \int d^2 \br \hat{\Psi}^\dagger_{\zeta}(\br) H_{\rm D-BdG}  \hat{\Psi}_{\zeta}(\br)~,
\end{equation}
where $\zeta=\pm$ denotes the sum over the valley indices and 
\begin{equation}
H_{\rm D-BdG} = \tau_z 
\left[  U(\br) \openone_\sigma + \frac{\hbar \vD}{i} (\partial_x \sigma_x + \partial_y \sigma_y)    \right]  
+ \tau_x \openone_\sigma \Real \Delta(\br)-\tau_y \openone_\sigma  \Imag\Delta(\br)~,
\end{equation}
\begin{equation}
\hat{\Psi}_{+}(\br) =  [\hat{\psi}_{A,  {\bm K},\uparrow}^\dagger(\br),\hat{\psi}_{B,  {\bm K},\uparrow}^\dagger(\br), \hat{\psi}_{A,  {\bm K}^\prime,\downarrow}(\br),\hat{\psi}_{B,   {\bm K}^\prime,\downarrow}(\br)]^\dagger~,
\end{equation}
\begin{equation}
\hat{\Psi}_{-}(\br) =  [-\hat{\psi}_{B, {\bm K}^\prime,\uparrow}^\dagger(\br),\hat{\psi}_{A,{\bm K}^\prime,\uparrow}^\dagger(\br), -\hat{\psi}_{B,{\bm K},\downarrow}(\br),\hat{\psi}_{A, {\bm K},\downarrow}(\br)]^\dagger~,
\end{equation}
$\vD \sim c/300$ is the Fermi velocity in monolayer graphene ($c$ is the speed of light), the identity $\openone_\sigma$ and the set of Pauli matrices $\{\sigma_x,\sigma_y,\sigma_z\}$ 
act on the ($A$ and $B$) sublattice subspace. The identity $\openone_\tau$ and $\{\tau_x,\tau_y,\tau_z\}$ act on the electron-hole pair subspace.
We approximate the superconductive order parameter and the scalar potential by a step-like profile, i.e.
\begin{equation}
\Delta(\br)= \Theta(|x|-L/2) \Delta_0 e^{i \phi_0(x)}~, 
\end{equation}
\begin{equation}
\phi_0(x)=\Theta(x) \phi_{\rm R} +\Theta(-x) \phi_{\rm L}~,
\end{equation}
\begin{equation}
U(\br)=-\mu_0\Theta(L/2-|x|)  -U_0\Theta(|x|-L/2)~,
\end{equation}
where  $\Theta(x)$ is the Heaviside step function, and $U_0 \gg |\mu_0|$.

{\bf Andreev Bound States.}
We are interested in the short junction limit $W\ll \xi \sim \hbar \vD/\Delta_0$, where $\xi$ is the superconducting coherence length.
In general, the spectrum of the D-BdG Hamiltonian consists of  Andreev bound states (ABSs) and a continuum of eigenstates.
The ABSs are subgap eigenstates, $|E|<\Delta_0$, and they are sensitive to the phase difference between the superconductive sides, $\phi=\phi_{\rm R}-\phi_{\rm L}$. 
They are spatially localized in the central normal phase region, while in the superconductive regions an evanescent tail is present.
On the other hand, eigenstates corresponding to the continuum spectrum with eigenergies above the gap, $|E|>\Delta_0$,  are spatially delocalized along the entire device~\cite{levchenko_prb_2006,samuelsson_prb_2000}. 
In the short junction limit, eigenstates with energies above the gap  do not depend on the phase difference $\phi$,  thus only ABSs carry the Josephson equilibrium supercurrent.
In this work, we neglect the continuum, focusing on the low-energy properties of the GJJs.
We project the D-BdG Hamiltonian $\hat{\cal H}_{\rm D-BdG}$ onto the subspace spanned by the ABSs by the projector $\hat{\cal P}_{\rm A}$, defining the Andreev Hamiltonian as $\hat{\cal H}_{\rm A}= \hat{\cal P}_{\rm A} \hat{\cal H}_{\rm D-BdG} \hat{\cal P}_{\rm A}$.
For a given value of the phase difference $\phi$, we express the Andreev Hamiltonian as
\begin{equation}\label{eqn:HA}
\hat{\cal H}_{\rm A}=\sum_{\zeta=\pm} \sum_k   \epsilon(k,\phi) \hat{\Sigma}^z_{\zeta,k}~,
\end{equation}
where $\hat{\Sigma}^z_{\zeta,k}=\hat{\gamma}_{+,\zeta,k}^\dagger \hat{\gamma}_{+,\zeta,k}-\hat{\gamma}_{-,\zeta,k}^\dagger \hat{\gamma}_{-,\zeta,k}$,
$\hat{\gamma}_{j,\zeta,k}$ represents the fermionic ABS operator labeled by
the subband index $j=\pm$ which denotes if the eigenenergy is below or above the Fermi level,
the valley index $\zeta=\pm$, and the $y-$component of the momentum $k$, that is a conserved quantity because the GJJ is invariant along the $y$ direction.
The ABSs of the subband which lays below (above) the Fermi level are called lower (upper) ABSs.
Each pair of valley index $\zeta$ and momentum $k$ identifies a two-level system with energy splitting $2  \epsilon(k,\phi)$, independent of the valley index and given by
\begin{equation}\label{eqn:epskphi}
 \epsilon(k,\phi)=\Delta_0 \sqrt{1-\tau(k) \sin^2(\phi/2)}~,
\end{equation}
where $\tau(k)=(\kF^2-k^2)/[\kF^2-k^2 \cos ^2 (L \sqrt{\kF^2-k^2} )]$ is the normal state transmission probability, and $\kF=\mu_0/(\hbar \vD)$ is the Fermi wavenumber~\cite{titov_prb_2006}. 
Within the subspace spanned by the ABSs, we express the Andreev current operator as
\begin{equation}\label{eqn:IA}
\hat{ I}_{\rm A}=
-
\frac{e \Delta_0^2}{\hbar} \sum_{\zeta=\pm} \sum_k \frac{\tau(k)}{ \epsilon(k,\phi)} \sin(\phi/2)  [\cos(\phi/2) \hat{\Sigma}^z_{\zeta,k} - \sqrt{1-\tau(k)} \sin(\phi/2) \hat{\Sigma}^x_{\zeta,k}]~,
\end{equation}
where the operators $\hat{\Sigma}^z_{\zeta,k}$  and
 $\hat{\Sigma}^x_{\zeta,k}=\hat{\gamma}_{+,\zeta,k}^\dagger \hat{\gamma}_{-,\zeta,k}+\hat{\gamma}_{-,\zeta,k}^\dagger \hat{\gamma}_{+,\zeta,k}$ are respectively diagonal and off-diagonal in the subband index $j$.
 The diagonal term is related to the supercurrents sustained by the respective ABSs,  while the off-diagonal term is mainly responsible for current fuctuations~\cite{zazunov_prl_2003}.
We note that the supercurrent is suppressed
in case of total reflection $\tau(k)\to 0$, and the off-diagonal  matrix elements of $\hat{ I}_{\rm A}$ 
become negligible for total transmission $\tau(k)\to 1$.
(See Supplementary Note 1 for the wavefunctions solving the stationary D-BdG equation for the subgap ABSs and derivation of
$\hat{ I}_{\rm A}$)

{\bf Dilute impurities.}
We model the dilute ensemble of impurities by the Anderson model~\cite{anderson_prev_1961}, which has been conveniently applied to describe the effect of disorder in other graphene based devices~\cite{farjam_prb_2011,yuan_prb_2010,wehling_prl_2010,skrypnyk_jpcondmat_2013}.
To start with, we consider $N_{\rm D}$ identical impurities, which respect the time reversal symmetry
\begin{equation}
\hat{\cal H}_{\rm D}=\sum^{N_{\rm D}}_{d=1} \hat{\Phi}^\dagger_d   \epsilon_0 \tau_z  \hat{\Phi}_d~, 
\end{equation}
where $\hat{\Phi}_d=[\hat{c}^\dagger_{d,\uparrow},\hat{c}_{d,\downarrow}]^\dagger$.
The electron tunneling between Andreev states and impurities states is expressed by a
potential $\hat{\cal V}_{\rm D}= \hat{\cal V}+ \hat{\cal V}^\dagger$
of the following general form (see Supplementary Note 2)
\begin{equation}
\hat{\cal V}=
\sum^{N_{\rm D}}_{d=1} \sum_{\zeta=\pm}  \int d^2 {\bm r} 
\hat{\Phi}^\dagger_d V_{d,\zeta}({\bm r}) \hat{\Psi}_\zeta(\bm r)~,
\end{equation}
and
\begin{equation}\label{eqn:Vdp}
  V_{d,+}({\bm r})=
  \begin{bmatrix}
v_{A,d} (\br)&v_{B,d}(\br)&0&0 \\
0&0&-v^\ast_{A,d} (\br)&-v^\ast_{B,d}(\br)
\end{bmatrix}~,
\end{equation}
\begin{equation}\label{eqn:Vdm}
  V_{d,-}({\bm r})=
  \begin{bmatrix}
-v^\ast_{B,d} (\br)&v^\ast_{A,d}(\br)&0&0 \\
0&0&v_{B,d} (\br)&-v_{A,d}(\br)
\end{bmatrix}~.
\end{equation}
The complete Hamiltonian of ABSs and impurities can be written in compact form by the following block decomposition
\begin{equation}\label{eqn:Htotblocks}
\hat{\cal H}_{\rm tot} = 
\begin{bmatrix}
\hat{\cal H}_{\rm A} &  \hat{\cal P}_{\rm A} \hat{\cal V}^\dagger  \\ 
\hat{\cal V} \hat{\cal P}_{\rm A} &  \hat{\cal H}_{\rm D} 
\end{bmatrix}~.
\end{equation}
We emphasize that the diagonal blocks $\hat{\cal H}_{\rm A}$ and $\hat{\cal H}_{\rm D} $ act onto two different subspaces, 
the ABSs and impurities subspaces respectively. 
The off-diagonal blocks connect the two subspaces.
The effect of disorder enters in the Green's function
\begin{equation}
\hat{\cal G}_{\rm tot}(\Omega) =(\Omega-\hat{\cal H}_{\rm tot})^{-1}~.
\end{equation}
By exploiting the block decomposition of the total Hamiltonian  in Eq.~\eqref{eqn:Htotblocks}, it is easy to express 
the block of the ABS's Green's function as follows
\begin{equation}\label{eqn:Gabs}
\hat{\cal G}(\Omega)=\hat{{\cal P}}_{\rm A}\hat{\cal G}_{\rm tot}(\Omega)\hat{{\cal P}}_{\rm A}=[\Omega-\hat{\cal H}_{\rm eff}]^{-1}~,
\end{equation}
where the effective Hamiltonian including the disordered ensemble of impurities reads
\begin{equation}
\hat{\cal H}_{\rm eff}=\hat{\cal H}_{\rm A}+\hat{\cal P}_{\rm A}\hat{\cal V}^\dagger(\Omega-\hat{\cal H}_{\rm D})^{-1}\hat{\cal V} \hat{\cal P}_{\rm A}~,
\end{equation}
and
\begin{equation}\label{eqn:VG0V_gen}
 \hat{\cal V}^\dagger(\Omega-\hat{\cal H}_{\rm D})^{-1}\hat{\cal V}=
 \int d^2 \br \int d^2 \br^\prime \sum_{\zeta,\zeta^\prime}
 \sum^{N_{\rm D}}_{d=1} \hat{\Psi}^\dagger_\zeta(\br) V_{d,\zeta}^\dagger(\br)
\Big( \frac{\Omega}{\Omega^2-\epsilon_0^2} \openone_\tau + \frac{\epsilon_0}{\Omega^2-\epsilon_0^2} \tau_z  \Big) 
 V_{d,\zeta^\prime}(\br^\prime) \hat{\Psi}_{\zeta^\prime}(\br^\prime)
 ~,
\end{equation}
for details see Supplementary Note 3.

Starting from a tight-binding description, and assuming that a generic impurity placed at $\br_d$ in correspondence of a carbon site, acts on the electron system in graphene at atomic scale~\cite{pellegrino_prb_2009}, the matrix elements of the short-range interaction potential, which appear in Eqs.~\eqref{eqn:Vdp}-\eqref{eqn:Vdm}, read
\begin{equation}\label{eqn:vtraps}
 v_{\alpha,d}(\br)=t_0 \sqrt{A_c}[m_d \delta_{\alpha,A}+(1-m_d)\delta_{\alpha,B}] e^{-i \frac{2\pi}{3} n_{ d}} \delta(\br - \br_d)~,
\end{equation}
where $t_0$ is a tunneling amplitude, $m_d$ ($n_d$) is an index taking the values $\{0,1\}$ ($\{-1,0,1\}$), and $A_{\rm c}=3\sqrt{3}a^2/2$ is the area of a unit cell~\cite{katsnelson_book_2012}.
The $m_d$ index is related to the presence of the $A/B$ sublattices, while $n_d$ index is a consequence of the hexagonal symmetry of the lattice (technical details on the microscopic treatment of the impurities are  in Supplementary Note 2). 
We assume a random distribution of the impurities positions $\br_d$, and of the indices ($m_d$, $n_d$), this justifies the  approximation of  homogeneity  $\sum_{d}\approx [N_D/(12 L_x W)] \sum_{m_d} \sum_{n_d} \int d^2 \br_d$, which gives
\begin{equation}\label{eqn:VG0V}
 \hat{\cal V}^\dagger(\Omega-\hat{\cal H}_{\rm D})^{-1}\hat{\cal V}=\frac{n_{\rm D} t_0^2}{2} \sum_{\zeta=\pm}
 \int d^2 \br  \hat{\Psi}^\dagger_{\zeta}(\br)  \Big( \frac{\Omega}{\Omega^2-\epsilon_0^2} \openone_\tau + \frac{\epsilon_0}{\Omega^2-\epsilon_0^2} \tau_z  \Big) \hat{\Psi}_{\zeta}(\br)~,
\end{equation}
where $n_{\rm D} =N_{\rm D}/N$, and $N=2WL_x/A_{\rm c}$.
By projecting this effective potential, Eq.~\eqref{eqn:VG0V}, onto the subspace spanned by the ABSs we obtain
 \begin{equation}
 \hat{\cal P}_{\rm A}\hat{\cal V}^\dagger(\Omega-\hat{\cal H}_{\rm D})^{-1}\hat{\cal V} \hat{\cal P}_{\rm A}= \frac{n_{\rm D} t_0^2 \Omega}{2(\Omega^2-\epsilon_0^2)} \sum_{\zeta=\pm}\sum_k \hat{\Sigma}^0_{\zeta,k}~,
 \end{equation}
where $\hat{\Sigma}^0_{\zeta,k}=\hat{\gamma}_{+,\zeta,k}^\dagger \hat{\gamma}_{+,\zeta,k}+\hat{\gamma}_{-,\zeta,k}^\dagger \hat{\gamma}_{-,\zeta,k}$.
Note that, if Anderson impurities are replaced with defects described by localized electrostatic $\delta$-potentials~\cite{skrypnyk_prb_2011}, within the homogeneity approximation, one finds a potential of the form $\propto \sum_{\zeta=\pm}
 \int d^2 \br  \hat{\Psi}^\dagger_{\zeta}(\br) \tau_z  \hat{\Psi}_{\zeta}(\br)$. Projecting this potential onto the
subspace spanned by the ABSs, one obtains that these defects have no effect on the ABSs.

If, instead of identical  Anderson impurities, we consider a set of impurities with a distribution of energies 
$ \rho_{\rm imp}(\epsilon)= \sum_l (N_{{\rm D},l}/N_{\rm imp})\delta(\epsilon-\epsilon_l)$, where $N_{\rm imp}=\sum_l N_{{\rm D},l}$ is the total number of impurities, the effective Andreev Hamiltonian takes the form
\begin{equation}\label{eqn:Heff}
\hat{\cal H}_{\rm eff}=\hat{\cal H}_{\rm A}+\frac{n_{\rm imp} t_0^2 u(\Omega)}{2}   
\sum_{\zeta=\pm} \sum_k \hat{\Sigma}^0_{\zeta,k} ~, 
\end{equation}
where $u(\Omega)=  \Omega \int d \epsilon \rho_{\rm imp}(\epsilon)/(\Omega^2-\epsilon^2)$, and $n_{\rm imp}=N_{\rm imp}/N$. Here, for simplicity the tunneling amplitude $t_0$ between ABSs and all types of impurities is approximated by a constant, independent of the type of impurity.
We emphasize that, due to the symmetries of the Hamiltonian, each pair of ABSs hybridizes independently with impurities states and the short-range interaction does not induce mixing of the upper and lower ABSs. In the following sections we will investigate how the spectral features of the entangled system enter the equilibrium supercurrent and its fluctuations.

\begin{figure}[ht]
\includegraphics[width=0.99\columnwidth]{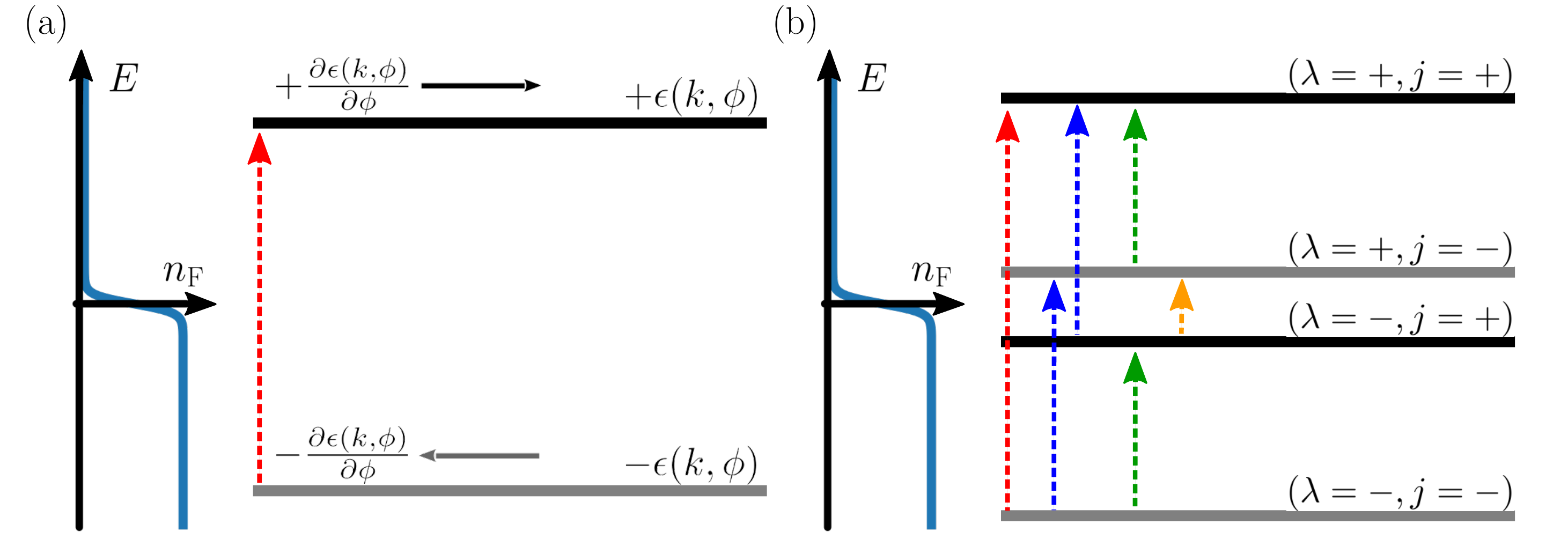}
\caption{
Scheme of the subgap levels for a generic $y$-component of the wavevector $k$, and superconductive phase difference $\phi$.
Panel a) refers to the clean limit ($\gamma\to \infty$), where the effect of the impurities  vanishes.
Gray (black) level represent the lower (upper) Andreev bound state (ABS), the gray (black) horizontal arrow sketches the direction of the supercurrent contribution $\propto -\partial \epsilon(k,\phi)/\partial \phi$ ($\propto \partial \epsilon(k,\phi)/\partial \phi$), and the red vertical dashed arrow denotes the possible transition.
On the left side, there is the Fermi-Dirac distribution at low temperature, $k_{\rm B} T \ll \Delta_0$, which shows that lower (upper) ABS is occupied (empty).
Panel b) refers to the limit $\gamma\to 0^+$.
Gray (black) levels $\Omega_{\lambda,-}$ ($\Omega_{\lambda,+}$) are associated to states which have a finite overlap on the lower (upper) ABSs of the clean graphene Josephson Junction (GJJ), and zero overlap on the upper (lower) ABSs of the clean GJJ. 
The colored vertical arrows represent the possible transitions between two subgap levels, arrows represented with the same color correspond to the same transition energy.
On the left side, there is the Fermi-Dirac distribution at low temperature, by comparing this with the vertical transitions, one can infer
that the transitions between ABSs labeled with opposite $j$ indices and identical $\lambda$ indices (green dashed lines) are suppressed by the Pauli blocking.
\label{fig:levs}}
\end{figure}

\begin{figure}[ht]
\centering
\includegraphics[width=0.99\columnwidth]{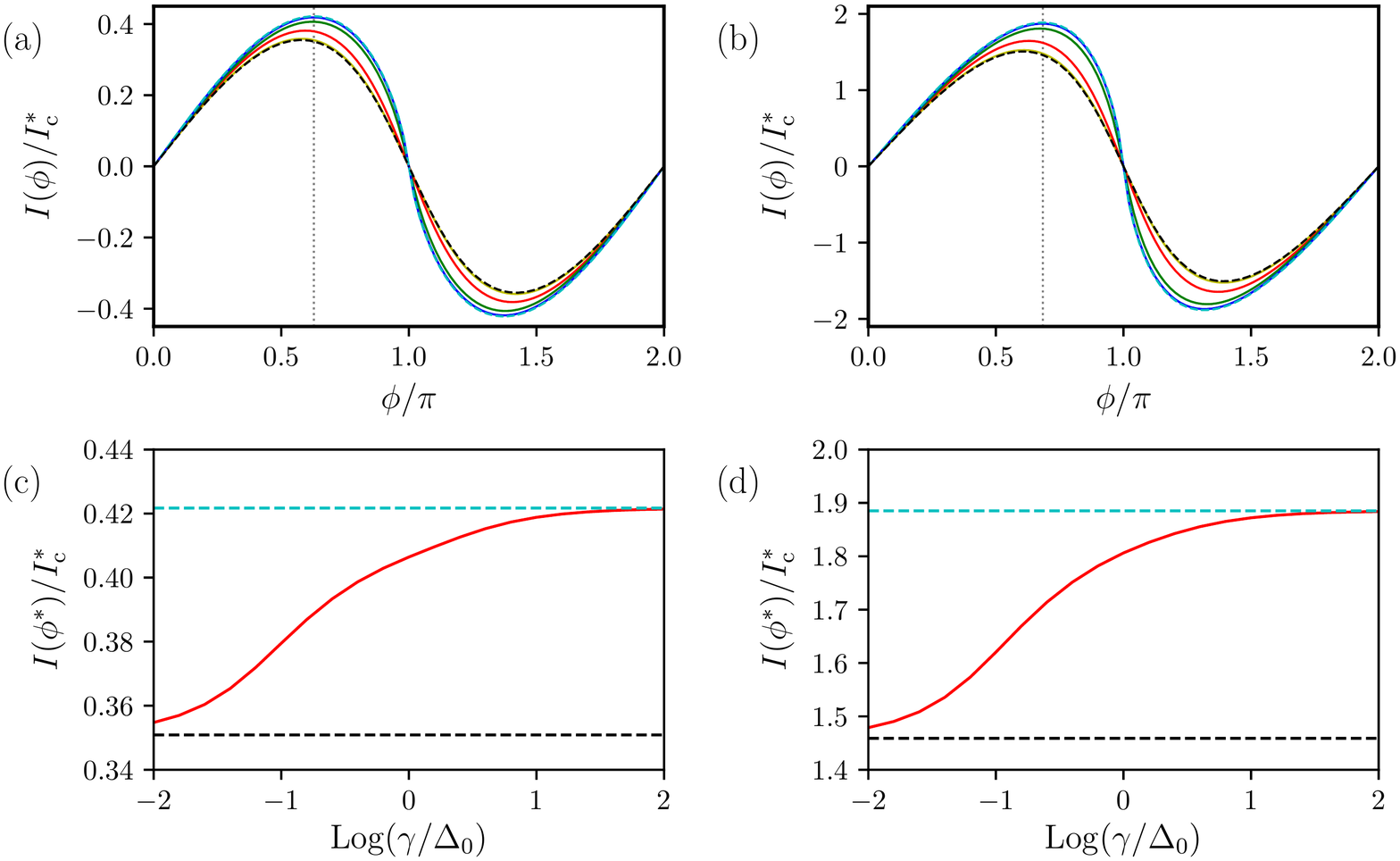}
\vspace{-2em}
\caption{Current-phase relation at zero temperature. Here, the impurity density is set at $n_{\rm imp} t_0^2 /\Delta_0^2=0.1$.
Panels a) and b) show the supercurrent as a function of the phase $\phi$  in units of $I_{\rm c}^\ast=e \Delta_0 W/(\hbar L)$, and the Fermi energy is set at $\mu_0=0$ and $\mu_0=5 \hbar \vD/L$, respectively.
In both panels one has  
$\gamma\to0^+$ (black dashed lines)  $\gamma=10^{-2} \Delta_0$ (yellow solid lines),  $\gamma=10^{-1} \Delta_0$ (red solid lines),  $\gamma= \Delta_0$ (green solid lines), 
$\gamma=10 \Delta_0$ (blue solid lines), $\gamma\to\infty$ (cyan dashed lines, i.e. the clean limit).
The dotted gray vertical line denotes $\phi^\ast$, which is the superconductive phase difference  such that $I(\phi^\ast)=\max_{\phi }I(\phi)$ in the clean GJJ, in particular $\phi^\ast=0.63\pi$ for $\mu_0=0$, and $\phi^\ast=0.68\pi$ for $\mu_0=5 \hbar \vD/L$.
Panels c) and d) show supercurrent $I(\phi^\ast)$ (red solid line), at zero temperature, as a function of $\gamma$, and the Fermi energy is set at $\mu_0=0$ and $\mu_0=5 \hbar \vD/L$, respectively.
In panels c) and d), the horizontal lines refer to two limiting cases: $I(\phi^\ast)$, at zero temperature, in the clean limit (horizontal cyan dashed line)
 and in the presence of single-energy impurities (horizontal black dashed line), i.e. $\gamma\to0^+$.
\label{fig:CPR}}
\end{figure}

{\bf Equilibrium supercurrent.}
The equilibrium supercurrent sustained by the GJJ in the short junction regime in the presence of 
a dilute distribution of impurities takes the following form 
\begin{eqnarray}\label{eqn:CPR}
 I(\phi)=\braket{\hat{I}_{\rm A}}=
 -
\frac{4e}{\hbar}   
\int  \frac{d \Omega}{2 \pi}  \sum_{j=\pm}   \sum_k j
\frac{\partial \epsilon(k,\phi)}{\partial \phi} n_F(\Omega) 
 A(j,k,\Omega)~,
\end{eqnarray}
where $ n_F(\Omega)=\{1+\exp[\Omega/(k_{\rm B} T)]\}^{-1}$~, and
\begin{eqnarray}
A(j,k,\Omega)&=&-2 {\rm Im}
\braket{j,\zeta,k|\hat{\cal G}_{\rm tot}(\Omega+i0^+)|j,\zeta,k
}=\braket{j,\zeta,k|\hat{\cal G}(\Omega+i0^+)|j,\zeta,k
}
\nonumber\\
&=&-2 {\rm Im} [\Omega+i0^+-j \epsilon(k,\phi)-n_{\rm imp} t_0^2 u(\Omega+i0^+)]^{-1}
\label{eq:spectralf}
\end{eqnarray}
is the spectral function, the last term accounts for coupling to the impurities. In the following, we will
consider a Lorentzian distribution of their energies $\rho_{\rm imp}(\epsilon)=(\gamma/\pi)/[(\epsilon-\epsilon_0)^2+\gamma^2]$, which gives $u( \Omega)=( \Omega+i \gamma)/[( \Omega+i \gamma)^2-\epsilon_0^2]$.
There is no dependence on the valley index $\zeta$ which introduces a degeneracy factor $2$ (details of the equilibrium Green's functions formalism are given in Supplementary Note 4).
In the CPR, the subband index $j$ in front of ABSs eigenenergies $\partial \epsilon(k,\phi)/\partial \phi$ is 
responsible for the opposite directions of the supercurrent carried by the two ABSs of each pair, for any value
of the  $y$-component of the wavevector.
For sake of simplicity, in the following discussion we set the central energy of the impurities at the Fermi energy, i.e. $\epsilon_0=0$. 
Under this condition the spectral function reads
\begin{eqnarray}
A(j,k,\Omega)&=&-2{\rm Im}\left[\Omega+i0^+-j \varepsilon(k,\phi) - \frac{n_{\rm imp} t_0^2}{\Omega+i\gamma}\right]^{-1}\\
&=& -2{\rm Im}\Bigg[ \frac{1}{\Omega-\Omega_{+,j}(k,\phi)}\frac{\Omega_{+,j}(k,\phi)+i\gamma}{\Omega_{+,j}(k,\phi)-\Omega_{-,j}(k,\phi)}+\frac{1}{\Omega-\Omega_{-,j}(k,\phi)}\nonumber\\
&\times&\frac{\Omega_{-,j}(k,\phi)+i\gamma}{\Omega_{-,j}(k,\phi)-\Omega_{+,j}(k,\phi)}
 \Bigg]~,\nonumber \,
\end{eqnarray}
which has two complex poles
\begin{equation}\label{eqn:Omegas}
\Omega_{\lambda,j}(k,\phi)=\frac{j \epsilon(k,\phi)-i\gamma}{2} +\lambda\sqrt{\left( \frac{j \epsilon(k,\phi)+i\gamma}{2} \right)^2 +\frac{n_{\rm imp} t_0^2}{2}}~,
\end{equation}
with $\lambda=\pm$. Symmetry properties and dependence of these complex energies on the system's physical parameters influence fundamentally the CPR. Here we discuss these properties in detail.
Since the system is electron-hole symmetric,  the poles in Eq.~\eqref{eqn:Omegas} have the following properties ${\rm Re}\Omega_{\lambda,j}(k,\phi)=-{\rm Re}\Omega_{-\lambda,-j}(k,\phi)$, and ${\rm sgn}[{\rm Re}\Omega_{\lambda,j}(k,\phi)]=\lambda$. 
In addition, $\Omega_{\lambda,j}(k,\phi)$ are even function of $k$, since the $k$ dependence originates from the transmission probability $\tau(k)$, see Eqs.~\eqref{eqn:Omegas} and~\eqref{eqn:epskphi}.
The dependence on the impurities parameters, in the dilute regime $n_{\rm imp} t_0^2/\Delta_0^2\ll1$, is as follows.
The two poles ${\rm Re}\Omega_{-,-}(k,\phi)$ and ${\rm Re}\Omega_{+,+}(k,\phi)$ are close to energies $- \epsilon(k,\phi)$ and  $+\epsilon(k,\phi)$  of the 
ABS of the clean GJJ, for any value of the doping level $\mu_0$.
Instead, ${\rm Re}\Omega_{+,-}(k,\phi)$ and ${\rm Re}\Omega_{-,+}(k,\phi)$ are close to the central energy $\epsilon_0$ of the impurities energy distribution which we have fixed at the Fermi energy.
The width of the impurities energies distribution, $\gamma > 0$, determines the finite lifetime of the resonances
at ${\rm Re}\Omega_{\lambda,j}(k,\phi)$.
For any $\gamma$, the hybridization between the ABSs and the impurity states is stronger in correspondence of the component $k$ such that $\tau(k) \sim 1$. Indeed, in proximity of the total transmission, the dispersion relation $\epsilon(k,\phi)$ moves close to the Fermi energy, where the distribution $\rho_{\rm imp}(\epsilon)$ is centered. 
In the limiting case $\gamma \to \infty$, for given $j$ and $k$, the spectral function tends to a single Dirac delta function at the ABSs energies, i.e. $A(j,k,\Omega) \to- 2 \pi \delta(\Omega-j\epsilon(k,\phi))$, corresponding to the clean GJJ.
Fig.~\ref{fig:levs}~a) sketches a couple of subgap levels $\pm \epsilon(k,\phi)$ for a generic $y$-component of the wavevector $k$ and superconductive phase difference $\phi$, in the clean limit ($\gamma \to \infty$).
Gray (black) level represents the lower (upper) ABSs, gray (black) horizontal arrow indicates the direction of the corresponding supercurrent contributions. 
The Fermi-Dirac distribution on the left-hand side of Fig.~\ref{fig:levs}~a) evidences that at low temperatures, $k_{\rm B} T \ll \Delta_0$, only the  lower ABS is occupied thus only its supercurrent contribution is active.
 In the opposite limit $\gamma\to0^+$, the poles in Eq.~\eqref{eqn:Omegas} reduce to the exact eigenenergies of the total Hamiltonian $\hat{\cal H}_{\rm tot}$, namely
$\Omega_{\lambda,j}(k,\phi) \to j \epsilon(k,\phi)/2 +  \lambda \sqrt{ \epsilon(k,\phi)^2/4 +n_{\rm imp} t_0^2/2}$.
The poles labeled by $\lambda=-$ ($\lambda=+$) lay energetically below (above) the Fermi energy.
For given $j$ and $k$, the spectral function becomes a weighted sum of two Dirac delta functions centered at those eigenenergies, $A(j,k,\Omega) \to -2\pi \sum_{\lambda=\pm} \delta(\Omega-\Omega_{\lambda,j}(k,\phi)) \Omega_{\lambda,j}(k,\phi)/[ \Omega_{\lambda,j}(k,\phi)- \Omega_{-\lambda,j}(k,\phi)]$.
Fig.~\ref{fig:levs}~b) shows the four subgap levels for $\gamma\to 0^+$, for a generic $y$-component of the wavevector $k$ and superconductive phase difference $\phi$. Here,  the states associated with the gray (black) levels $\Omega_{\lambda,-}$ ($\Omega_{\lambda,+}$)  labeled by the subband index $j=-$ ($j=+$) have a finite overlap on the lower (upper) 
and zero overlap on the upper (lower) ABSs of the clean GJJ,  and they carry a supercurrent contribution  $\propto -\partial \epsilon(k,\phi)/\partial \phi$ ($\propto \partial \epsilon(k,\phi)/\partial \phi$). 
By comparing the Fermi-Dirac distribution at low temperature, $k_{\rm B} T \ll \Delta_0$, with the structure of levels, one  sees that the occupied states are those labeled by $(\lambda=-,j=-)$ and $(\lambda=-,j=+)$. They have finite overlap with the lower and upper ABS respectively. Therefore, they carry  supercurrent contributions in opposite directions. In other words, the presence of impurities activates the supercurrent contribution of the upper ABS  also at zero temperature, reducing the total supercurrent contribution for each $k$. This compensating effect on the supercurrent is largest when the hybridization is maximal, namely for $k$ such that $\tau(k) \sim 1$. 
Figs.~\ref{fig:CPR}~a) and b) show the  CPR at zero temperature, obtained by using Eq.~(\ref{eqn:CPR}), with $n_{\rm imp} t_0^2/\Delta_0=0.1$, for different values of $\gamma$,  at zero doping, $\mu_0=0$ (Figs.~\ref{fig:CPR}~a)) and at the finite doping level $\mu_0=5\hbar \vD /L$ (Figs.~\ref{fig:CPR}~b)).
In both cases we observe that in the clean limit ($\gamma\to\infty$) 
the CPR shows the largest skewness and critical current. 
For finite values of  $\gamma$, the hybridization between the ABSs and the impurities sets in. As a consequence,
both  skewness and  critical current reduce.
This effect can explained by observing that the modes mainly affected by the presence of disorder are the high transparent ones, $\tau(k) \sim 1$.
These modes are also the ones largely responsible for skewness of the clean GJJ giving a supercurrent contribution
 $\propto {\partial \epsilon(k,\phi)}/{\partial \phi} \sim \Delta_0 \sin(\phi/2)$, whereas modes with low transparency 
 imply the standard sinusoidal dependence $\propto \sin(\phi)$. 
 The phase difference where the supercurrent is maximal $\phi_{\rm max}$, depends on $\gamma$ and on the doping level $\mu_0$.
In particular, we denote with $\phi^\ast$  the value in the clean limit  $\phi^\ast=\lim_{\gamma\to \infty}\phi_{\max}$.
The effect of the impurities energy distribution, $\gamma$, on the maximal supercurrent  is illustrated in  Figs.~\ref{fig:CPR}~c) and d)
where we plot the supercurrent evaluated at $\phi=\phi^\ast$, for two different values of the doping.
In both cases we observe a monotonic increase of the supercurrent with increasing $\gamma$, the supercurrent is minimal
for the Dirac delta energy distribution, i.e. $\gamma\to0^+$.

\begin{figure}[ht]
\includegraphics[width=0.99\columnwidth]{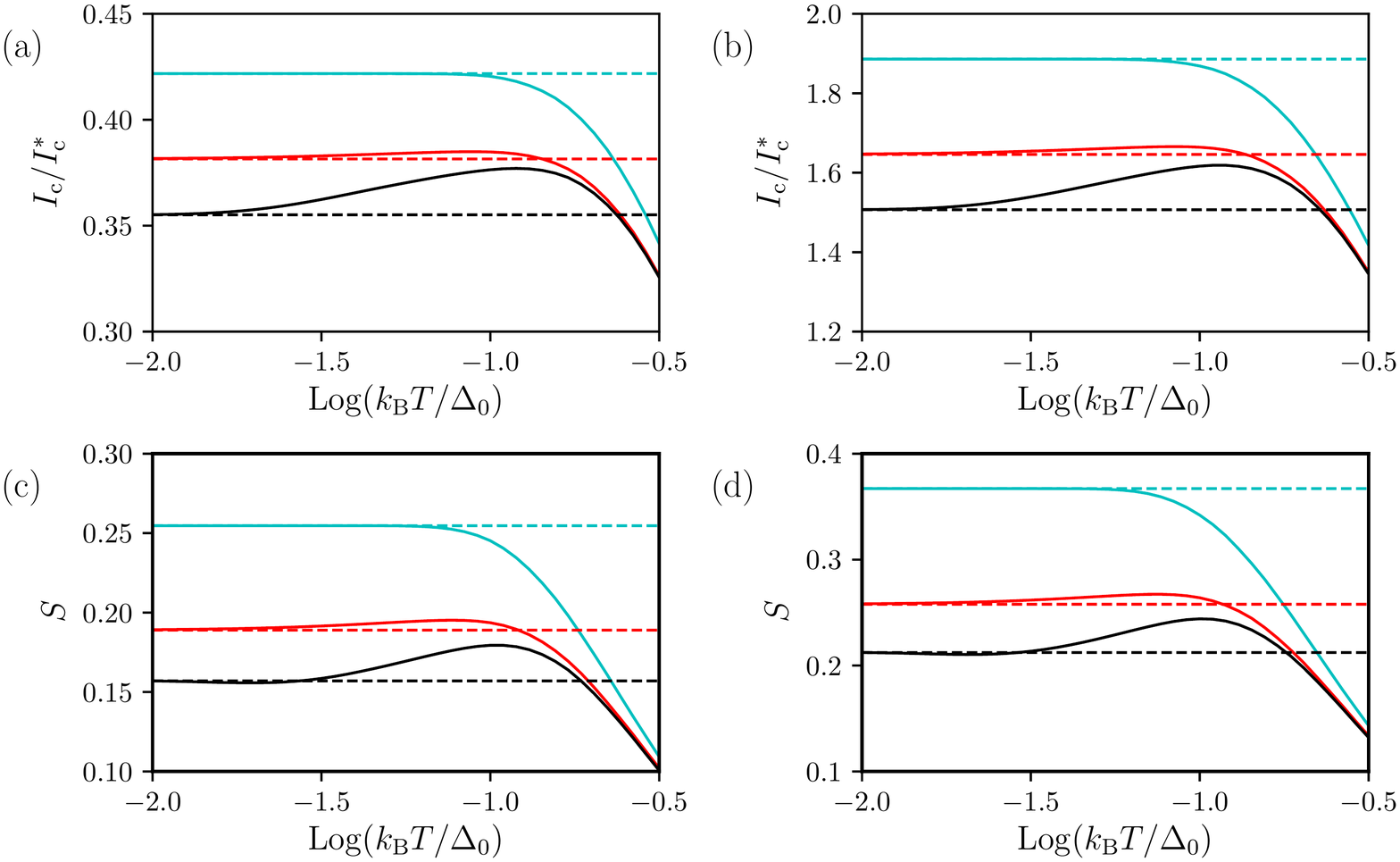}
\vspace{-2em}
\caption{
Critical current and skewness $S$ as a function of temperature (solid lines), compared with  the respective values at zero temperature (horizontal dashed lines).
Panels a) and b) represent the critical current,  in units of $I^\ast_{\rm c}$, and the Fermi energy is set at $\mu_0=0$ and $\mu_0=5 \hbar \vD/L$, respectively.
Panels c) and d) show the skewness, and the Fermi energy is set at $\mu_0=0$ and $\mu_0=5 \hbar \vD/L$, respectively.
 In all panels, one has $\gamma\to 0^+$ (black lines), $\gamma=10^{-1} \Delta_0$ (red lines), $\gamma\to\infty$ (cyan lines), the temperature dependence of the order parameter $\Delta_0$ is neglected, and the impurity density is set at $n_{\rm imp} t_0^2 /\Delta_0^2=0.1$.
\label{fig:Ic_Sk}}
\end{figure}

According to the D-BdG theory, the critical current in short ballistic GJJ decreases monotonically
with temperature~\cite{titov_prb_2006}. 
Whereas, this qualitative trend has been observed in recent experiments,  smaller
values of the critical current than one at zero temperature accompanied by unexplained irregularities have also been reported   ~\cite{borzenets_prl_2016,english_prb_2016,nanda_nanol_2017,park_prl_2018}. Similar discrepancies have been observed also
for the temperature dependence of the skewness \cite{nanda_nanol_2017}.  
Here, we discuss the temperature dependence of the critical current and skewness resulting from the hybridization of ABSs with impurities which provide an alternative mechanism for the reported deviations. 
Fig.~\ref{fig:Ic_Sk} shows the critical current, panels~a) and~b), and  the
skewness defined as $S=2 \phi_{\rm max}/\pi-1$, panels~c) and~d),
as a function of temperature (solid lines), compared with  the respective values at zero temperature (horizontal dashed lines).
Figs.~\ref{fig:Ic_Sk}~a) and~c) 
refer to the undoped case, while Figs.~\ref{fig:Ic_Sk}~b) and~d)
refer to the doped case with $\mu_0=5 \hbar \vD/L$. 
The temperature dependencies in the clean limit, $\gamma\to\infty$, derive form the thermal population of pairs of
ABSs carrying opposite supercurrents  inducing a monotonic decrease of the critical current with  temperature (cyan lines). Moreover, for any given phase difference $\phi$, thermal activation of the
upper Andreev levels is mainly effective for wavevector components $k$ corresponding to large transmission $\tau(k)$,
since the corresponding energies $\epsilon(k,\phi)$ are closer to the Fermi energy. These modes are also responsible 
for  the forward skewness of the CPR. Therefore, in the clean limit, $S$ diminishes with increasing temperature, as shown in Figs.~\ref{fig:Ic_Sk}~c) and~d) (cyan lines).
Instead the presence of single-energy impurities (black lines), i.e. $\gamma\to0^+$,  both the critical current and skewness display a nonmonotonic temperature dependence. This behavior can be understood considering the thermal population of hybridized Andreev-impurities energies sketched in Fig.~\ref{fig:levs}~b).
For small temperatures $k_{\rm B} T \lesssim n_{\rm imp} t_0^2/\Delta_0$ the only levels above the Fermi energy which 
become populated are levels $(\lambda=+,j=-)$.
They carry a supercurrent in the same direction of the dominant contribution due to the lowest hybridized level $(\lambda=-,j=-)$, while it is opposite to the contribution of the level $(\lambda=-,j=+)$.
In other words, the thermal activation of supercurrent contributions of the hybridized levels $(\lambda=+,j=-)$ suppresses the effect of the disorder and induces an increase both of the critical current and the forward-skewness.
For larger temperatures, such that $ n_{\rm imp} t_0^2/\Delta_0 < k_{\rm B} T < \Delta_0$, for each $k$, the supercurrent contributions of the levels  $(\lambda=-,j=+)$ and  $(\lambda=+,j=-)$ are comparable and cancel each other. 
On the other side the population of the topmost level  $(\lambda=+,j=+)$ becomes significant and contributes with a supercurrent
summing up to the one due to the hybridized ground state. 
As a consequence, the thermal trend becomes one observed in the clean limit (cyan lines). At these range of temperatures, critical current and skewness are decreasing.
Finally, the red line in Fig.~\ref{fig:Ic_Sk}  shows the temperature dependence of the critical current and the skewness in the presence of a  finite width $\gamma=\Delta_0/10$, which are qualitatively similar to the case with a single-energy (cyan lines), but the finite width $\gamma \sim n_{\rm imp} t_0^2/\Delta_0 $, makes the increasing dependencies of the temperature less visible.
Thus hybridization between ABSs and impurities originates smaller critical current and skewness than the clean limit expectation based on the BdG theory, but nonmonotonic temperature dependence.

\begin{figure}[ht]
\includegraphics[width=0.99\columnwidth]{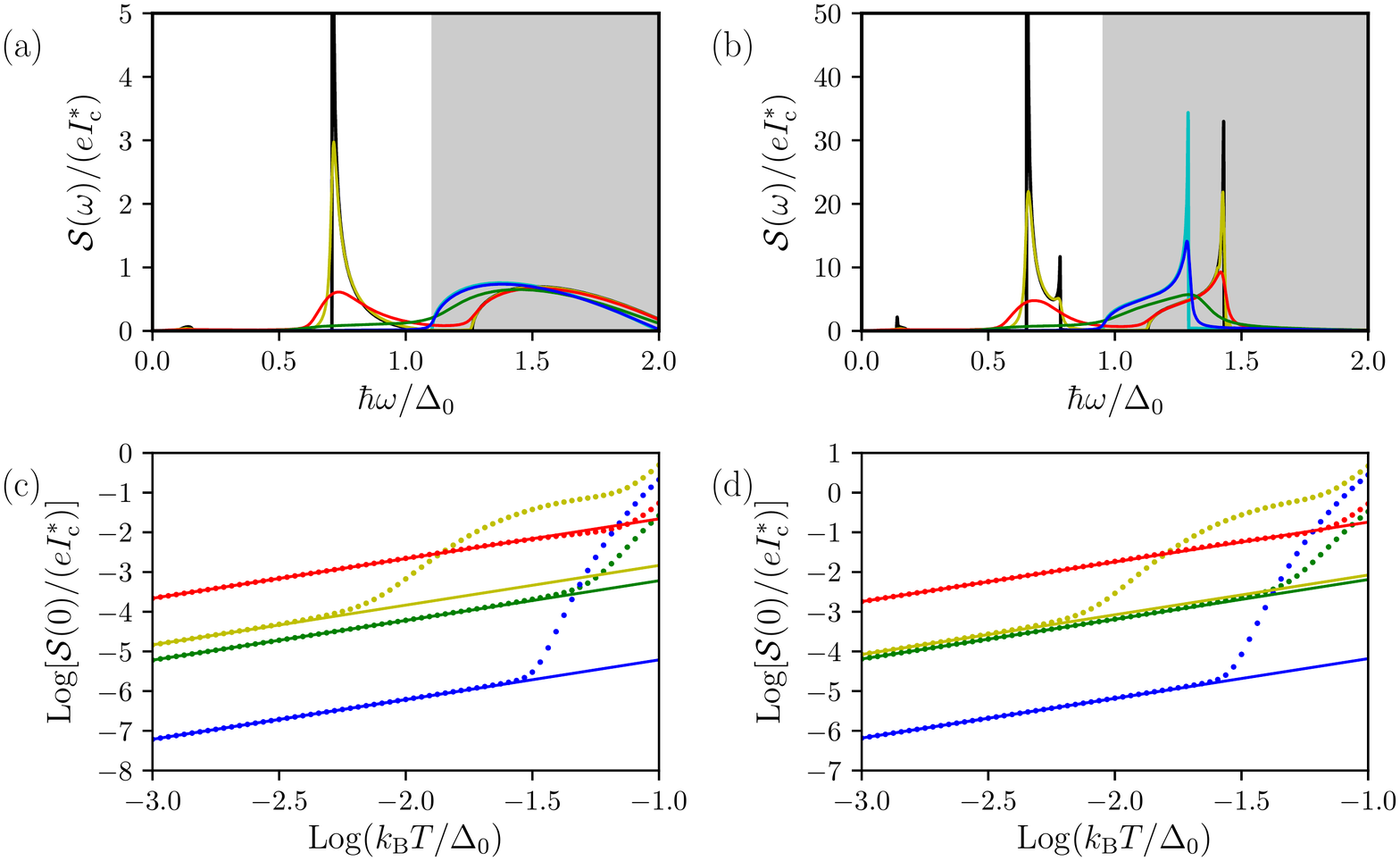}
\vspace{-2em}
\caption{
Supercurrent power spectrum.
Panels a) and b) show  $\cal S(\omega)$, in units of $e I^\ast_{\rm c}$ at  $\phi=\phi^\ast$, as a function of frequency $\omega$, and the Fermi energy is set at $\mu_0=0$ and $\mu_0=5 \hbar \vD/L$, respectively.
In both panels a) and b), one has $T=10^{-2} \Delta_0/k_{\rm B}$, $n_{\rm imp} t_0^2 /\Delta_0^2=0.1$,  $\gamma=0^+$ (black lines)  $\gamma=10^{-2} \Delta_0$ (yellow lines),  $\gamma=10^{-1} \Delta_0$ (red lines),  $\gamma= \Delta_0$ (green lines), $\gamma=10 \Delta_0$ (blue lines), and $\gamma=\infty$ (cyan lines).
The shaded region is the frequency domain where the supercurrent power spectrum is non-zero in a clean GJJ.
Panels c) and d) show the static supercurrent power spectrum ${\cal S}(0)$, in units of $e I^\ast_{\rm c}$ at  $\phi=\phi^\ast$, as a function of temperature, in a log-log scale, and the Fermi energy is set at $\mu_0=0$ and $\mu_0=5 \hbar \vD/L$, respectively.
In both panels c) and d), one has   
$\gamma=10^{-2} \Delta_0$ (yellow  circles),  $\gamma=10^{-1} \Delta_0$ (red circles),  $\gamma= \Delta_0$ (green  circles), 
$\gamma=10 \Delta_0$ (blue circles), each colored solid line represents the corresponding low temperature linear behaviour by Eq.~\eqref{eqn:staticS}. The temperature dependence of the order parameter $\Delta_0$ is neglected, and the impurity density is set at $n_{\rm imp} t_0^2 /\Delta_0^2=0.1$. 
\label{fig:S}}
\end{figure}

{\bf Supercurrent noise.}
A convenient quantity to identify spectral features of the hybridized system is the supercurrent noise
spectrum. As a difference with the CPR, which reflects the overall effect of the hybridized system, the supercurrent noise 
spectrum directly probes the possible absorption/emission frequencies because of the fluctuation-dissipation theorem ~\cite{giulianivignale_book}.
In particular, for $\omega>0$, ${\cal S}(\omega)$ gives the absorption spectrum.
Therefore, the supercurrent power spectrum can be used for a spectroscopic analysis of the source of disorder.
For fixed phase difference $\phi$, 
the equilibrium supercurrent fluctuations are expressed by noise power spectral density
\begin{equation}\label{eqn:Sdef}
 {\cal S}(\omega)=
 \int^\infty_{-\infty} dt e^{i \omega t}
[ \langle \hat{ I}_A(t)  \hat{ I}_A(0)\rangle-\langle \hat{ I}_A(t)\rangle \langle \hat{ I}_A(0)\rangle]~,
\end{equation}
where $\langle \cdots \rangle$ denotes the thermal equilibrium average of the entire system and the Andreev current operator  $\hat{ I}_A$  defined in Eq.~\eqref{eqn:IA}.
After algebraic manipulations (shown in detail in Supplementary Note 4), we obtain
\begin{eqnarray}\label{eqn:S_LRT}
{\cal S}(\omega)&=&   \hbar  \sum_{j=\pm,j^\prime=\pm} \sum_{\zeta=\pm} \sum_{k} \int \frac{d \Omega}{2 \pi} 
n_F(\Omega)[1-n_F(\Omega+\hbar \omega)] | \braket{j, \zeta, k |\hat{I}_{\rm A}|j^\prime ,\zeta, k } |^2\\
&\times& A(j,  k ,\Omega)  A(j^\prime,  k,\Omega+\hbar \omega) \, .\nonumber 
\end{eqnarray}
where the spectral function is given by Eq.~\eqref{eq:spectralf}.
Figs.~\ref{fig:S}~a) and b) show ${\cal S}(\omega)$ 
evaluated at $\phi=\phi^\ast$ for zero and finite doping and different widths $\gamma$ of the impurities energy distribution, at $T=10^{-2} \Delta_0/k_{\rm B}$. 
In the clean limit, spectral features are present only in the  
frequency domain  $2 \Delta_0 |\cos(\phi^\ast/2)|\le \hbar \omega\le 2 \Delta_0$ (gray shaded region).
The first qualitative feature of the presence of the dilute impurities is the appearance of additional spectral features at
smaller frequencies (white region).

The supercurrent power spectrum can be explained in terms of the transitions indicated in the scheme of Fig.~\ref{fig:levs}~b).
For a generic $y$-component of the wavevector $k$, there are four possible energies indicated by the colored dashed vertical arrows.
The transition with largest energy (red dashed  arrow) links the levels labeled by $(\lambda=-,j=-)$ and $(\lambda=+,j=+)$, the transition energy lays in the interval
$\Delta_0 < \hbar \omega < 2\Delta_0$. The two levels involved  collapse respectively to the lower and upper Andreev level
by turning off the interaction, $t_0\to 0$.
The transitions at intermediate energies, i.e. $\hbar \omega \lesssim \Delta_0$, 
can be classified in two types, the first one is $(\lambda=-,j)\to(\lambda=+,j)$ (blue dashed) and the second one $(\lambda,j=-)\to(\lambda,j=+)$ (green dashed).
The latter class of transitions  (green dashed) 
are strongly suppressed by the Pauli blocking. 
Finally, there is a class of very low energy transitions, $ \hbar \omega \ll \Delta_0$, (orange dashed) between the levels  $(\lambda=+,j=-)$ and $(\lambda=-,j=+)$. By turning off the interaction 
these two states have no overlap with the ABSs, so they do not contribute to the supercurrent.

In order to understand the origin of the main features of the supercurrent power spectrum, we first focus on the case with $\gamma\to0^+$. 
Here, for any generic $\phi$, the supercurrent power spectrum shows several square root divergences, each singularity occurs at an energy $\hbar \omega$ that corresponds to an extremum of the energy difference between the two subgap levels involved in the transition $\Omega_{\lambda,j}(k,\phi)-\Omega_{\lambda^\prime, j^\prime}(k,\phi)$.
Since $\partial \Omega_{\lambda,j}(k,\phi)/\partial k=(\partial \Omega_{\lambda,j}(k,\phi)/\partial \tau(k))(\partial \tau(k)/\partial k)$, the extrema of $\Omega_{\lambda,j}(k,\phi)-\Omega_{\lambda^\prime, j^\prime}(k,\phi)$ occur at the wavenumber $k$  where also the transmission probability $\tau(k)$ is extreme. 
In fact $\tau(k)$ is a bounded even function ($0\leq \tau(k) \leq 1$), which takes its maximum value $\tau(k)=1$ (total transmission), for $k=0$ (Klein tunneling) and for $k=\pm \sqrt{\kF^2-(\pi n)^2/L^2}$ (stationary wave condition). The number of wavevector component $k$ which fulfills the stationary wave condition is $2{\rm Int}(\kF L/\pi)$, and it depends on the doping level $\mu_0$.
In correspondence of the values of $k$ which give total transmission the energy differences are equal to 
$\Omega_{\lambda,j}(0,\phi)-\Omega_{\lambda^\prime, j^\prime}(0,\phi)$.
In between the $2{\rm Int}(\kF L/\pi)+1$ values of $k$ where $\tau(k)=1$, the transmission probability $\tau(k)$ takes $2{\rm Int}(\kF L/\pi)$  local minima for $k$ values solving the transcendental equation $\kF^2\sin(L \sqrt{\kF^2-k^2})=L k^2 \sqrt{\kF^2-k^2}\cos(L \sqrt{\kF^2-k^2})$, such that $|k|<\kF$.
By analyzing the supercurrent power spectrum ${\cal S}(\omega)$, we see that
if the two subgap levels  involved $(\lambda, j)$ and $(\lambda^\prime, j^\prime)$ are such that $ j^\prime=j$ then square root divergences may appear in correspondence of  both a global maximum and a local minimum of the transmission probability. %
Thus, there are ${\rm Int}(\kF L/\pi)$ square root divergences associated with the minima and a further square root divergence associated with total transmission.
On the other hand, if the two subgap levels involved are  $(\lambda, j)$ and $(\lambda^\prime, j^\prime)$ such that $ j^\prime=-j$ then the square root divergences appear in correspondence only of a local minimum of the transmission probability, since the matrix element $\braket{j,\zeta,k|\hat{I}_{\rm A}|-j,\zeta,k}$ vanishes for $k$ such that $\tau(k)=1$, independently of the valley index $\zeta$. Thus, for  $ j^\prime=-j$ there are ${\rm Int}(\kF L/\pi)$ square root divergences.
At a finite $\gamma>0$, the exact levels described above are replaced by resonances with a finite lifetime, see Eq.~\eqref{eqn:Omegas}.  The square root divergences become resonances and the supercurrent power spectrum is a regular function of the frequency. For $\gamma \to\infty$, the levels $(\lambda=-,j=-)$ and $(\lambda=+,j=+)$ collapse to the Andreev levels of the clean GJJ that have an infinite lifetime, whereas the levels $(\lambda=-,j=+)$ and  $(\lambda=+,j=-)$ become ill-defined resonances with a vanishing lifetime. For $\gamma \gg \Delta_0$, the supercurrent power spectrum tends to the profile of a clean GJJ (see cyan lines in Figs.~\ref{fig:S}~a) and b)), where there is no signal in the low frequency domain $\hbar \omega\lesssim \Delta_0$  (see white regions in Figs.~\ref{fig:S}~a) and b)).
In Fig.~\ref{fig:S}~a), which refers to the undoped case, one sees that for $\gamma\to0^+$ (black line)  the supercurrent power spectrum shows a single square root divergence placed at the intermediate energy $\hbar \omega=\sqrt{\Delta_0^2\cos^2(\phi^\ast)+2n_{\rm imp}t_0^2} \approx 0.71 \Delta_0$, while in the clean limit $\gamma\to \infty$ (cyan line) 
the supercurrent power spectrum is a smooth function.
The case of finite doping is shown in Fig.~\ref{fig:S}~b), where $\mu_0=5 \hbar \vD/L$. In the limiting case $\gamma\to0^+$ (black line)  the supercurrent power spectrum shows four square root divergences. In particular, there is a divergence in the shaded region  $\hbar \omega=\epsilon(\bar{k},\phi^\ast)+\sqrt{\epsilon(\bar{k},\phi^\ast)^2+2n_{\rm imp}t_0^2} \approx 1.43 \Delta_0$ (where the value $\bar{k}\approx2.9/L$ solves the transcendental equation shown above),  there are two divergences in the intermediate energies, i.e.  $\hbar \omega=\sqrt{\epsilon(\bar{k},\phi^\ast)^2+2n_{\rm imp}t_0^2} \approx 0.78 \Delta_0$ and $\hbar \omega=\sqrt{\Delta_0^2\cos^2(\phi^\ast)+2n_{\rm imp}t_0^2} \approx 0.71 \Delta_0$, and a further square root divergence appears at low energy $\hbar \omega=-\epsilon(\bar{k},\phi^\ast)+\sqrt{\epsilon(\bar{k},\phi^\ast)^2+2n_{\rm imp}t_0^2} \approx 0.14 \Delta_0$.
In the clean limit $\gamma\to \infty$ (cyan line), only the square root divergence in shaded region holds, and it is red-shifted at $\hbar \omega=2\epsilon(\bar{k},\phi^\ast)\approx 1.29 \Delta_0$.

 We note that, because of disorder, the zero frequency current noise reduces to the linear
thermal noise behaviour for sufficiently small temperatures. The slope of the linear dependence can be
related to the impurity energy distribution. 
Indeed, the limit $k_{\rm B} T\ll \gamma, \Delta_0$, one has
\begin{eqnarray}\label{eqn:staticS}
{\cal S}(0)&=&   \hbar  \sum_{j=\pm,j^\prime=\pm} \sum_{\zeta=\pm} \sum_{k} \int \frac{d \Omega}{2 \pi} 
\frac{1}{4 \cosh^2(\frac{\Omega}{2 k_{\rm B} T})} | \braket{j, \zeta, k |\hat{I}_{\rm A}|j^\prime ,\zeta, k } |^2\\
&\times& A(j,  k ,\Omega)  A(j^\prime,  k,\Omega)\nonumber \\
&\approx&   \left[ \frac{\hbar}{2 \pi} \sum_{j=\pm,j^\prime=\pm} \sum_{\zeta=\pm} \sum_{k} 
 | \braket{j, \zeta, k |\hat{I}_{\rm A}|j^\prime ,\zeta, k } |^2 A(j,  k ,0)  A(j^\prime,  k,0) \right]  k_{\rm B} T  \nonumber \\
&=&\sin^2(\phi/2)  \left\{ \frac{8 \pi e^2 \Delta_0^2}{\hbar }  \sum_k
\Bigg[
\frac{\tau(k)}{\pi} \frac{(n_{\rm imp} t_0^2)/(2\gamma)}{\epsilon^2(k,\phi)+(n_{\rm imp} t_0^2)^2/(2\gamma)^2} \Bigg]^2 \right\} k_{\rm B} T
~,\nonumber
\end{eqnarray}
where we have approximated $1/[4 \cosh^2(\frac{\Omega}{2 k_{\rm B} T})] \to k_{\rm B} T\delta(\Omega)$. 
We have assumed that any spectral function $A(j,  k ,\Omega) $ is smooth, thus it can be approximated as
 $A(j,  k ,0)=(n_{\rm imp} t_0^2/\gamma)/[\epsilon^2(k,\phi)+(n_{\rm imp} t_0^2)^2/(2\gamma)^2]$.
 Note that the slope of the linear temperature behavior depends on 
the width $\gamma$, in particular it vanishes in both limits $\gamma\to 0^+$ and $\gamma \to \infty$.
The dependence on $\gamma$ of ${\cal S}(0)$ is shown in Fig.~\ref{fig:S}~c) and Fig.~\ref{fig:S}~d) for zero and finite doping, respectively.

{\bf Conclusion} 

In this work we have investigated the modifications of the Andreev spectrum in a short ballistic GJJ due to the hybridization with a dilute set of non-magnetic impurities homogeneously distributed below the entire device.
The ABSs are described by a D-BdG model. Within this formalism, we considered a set of impurities described by the Anderson model, and with a Lorentzian distribution of energies about the Fermi energy with a width $\gamma$.
We remark that our analytic formalism can be readily applied also to other distributions of impurity energy levels.
Here, we have obtained that, both with undoped and doped normal region, for any value of the energy width $\gamma$,
the dilute ensemble of impurities causes a reduction of the critical current and, more prominently, of the skewness the current-phase relation.
In an impurity-free GJJ the current phase relation is skewed by very high transmittance channels~\cite{golubov_rmp_2004,titov_prb_2006}. Here, we found that exactly these ABSs, labeled by $k$ 
such that $\tau (k) \sim 1$,
are mainly hybridized with the impurity levels. This phenomenon leads to a reduction of the supercurrent contributions that induce the skewness of the CPR.
 Moreover, we found that thermal excitations can inhibit this mechanism due to the population of
 higher energy hybridized ABS-impurity states carrying opposite supercurrent.
 This determines a counterintuitive increase of both the critical current and the skewness around a range of low temperatures, such that $k_{\rm B} T \sim t_0^2 n_{\rm imp}/\Delta_0$.
Within our formalism, we have also derived the power spectrum of the supercurrent both with undoped and doped normal region.
This quantity turns out to be a powerful spectroscopic tool of the hybridized spectrum.
In particular, for an impurity-free GJJ, we find a low frequency domain, $0 \le \omega < 2 \Delta_0|\cos(\phi/2)|/\hbar$, where the power spectrum of the supercurrent is vanishing, and it is tunable by the superconductive phase difference $\phi$.
Because of the hybridization of the ABSs with impurity levels, resonances appear in the low frequency region whose position and number have been predicted. 
Moreover, we have connected all the peaks of the power spectrum to features of the transmittance probability $\tau(k)$. 
Finally, we have seen that at very low temperatures ($k_{\rm B} T \ll \Delta_0,\gamma$), the power spectrum of the supercurrent displays a linear dependence on the temperature, with a slope related to the spectral weight at the Fermi level, which vanishes both for $\gamma\to0^+$ and $\gamma \to \infty$.
These results highlight the extraordinary potentialities of the supercurrent in a GJJ and its equilibrium noise as probes of impurities accidentally present even in clean van der Waals heterostructures.  
Future work will be devoted to study the effect of Anderson impurities on GJJ in the long and intermediate junction limits, by taking into account the Andreev continuum which cannot be disregarded.


\vspace*{5mm}
\textbf{Methods}

The integration above has been performed with Python numerical routines, 
in particular we have used the free and open-source library Scipy~\cite{scipy}. 

\textbf{Data availability}

The data that support the findings of this study are available from the corresponding
author upon request.

\vspace*{5mm}


\bibliography{manuscript}
\vspace*{5mm}
\textbf{Acknowledgments.}
The authors thank G. G. N. Angilella, F. Bonasera, R. Fazio, P. Hakonen, and V. Varrica for illuminating
discussions and fruitful comments on various stages of this work. 
This research was supported by the Universit\`a degli
Studi di Catania, Piano di Incentivi per la Ricerca di
Ateneo 2020/2022 (progetto QUAPHENE and progetto Q-ICT), and   Centro Siciliano di Fisica Nucleare e Struttura della Materia (CSFNSM).
\vspace*{5mm}

\textbf{Author contributions.}
All the authors conceived the work, agreed on the approach to pursue, analysed and
discussed the results. F.M.D.P. performed the analytical and numerical calculations,
E.P. and G.F. supervised the work.\\

\textbf{Competing interests:} The authors declare no competing interests.

\newpage
\appendix
\numberwithin{equation}{section}

\section{Supplementary Note 1}

We are interested in the Andreev bound states (ABSs) of a short graphene Josephson Junction (GJJ) in the ballistic transport regime. 
This system is uniform  along the $y$-direction, and it is described in terms of the Dirac-Bogoliubov-De Gennes (D-BdG) approach in the main text.
We look for solutions of the stationary D-BdG equation of the factorized form
\begin{equation}\label{eqn:plwave}
{\bm \varphi}_{k,E,\zeta}(\br)= \frac{e^{i k y}}{\sqrt{W}}  \tilde{\bm \varphi}_{k,E}(x)~,
\end{equation}
where  $|E|<\Delta_0$ is the energy, $k$  is the $y$-component of the wavevector, and $\zeta=\pm$ is a valley index.
We remind that the D-BdG  Hamiltonian in Eq. \eqref{eq:HDBdG_CB}  is diagonal in the valley index $\zeta$, thus we  omit this degree of freedom in the spinor $\tilde{\bm \varphi}_{k,E}(x)$.
Along the  $x$-direction  we consider a sharp partition in 
superconducting-normal-superconducting sectors.  For each sector we look for a solution of the form \eqref{eqn:plwave}, thus
\begin{equation}
 \tilde{\bm \varphi}_{k,E}(x)=\Theta(-L/2-x)\tilde{\bm \varphi}^{(\rm S-left)}_{k,E}(x)+\Theta(L/2-|x|)\tilde{\bm \varphi}^{(\rm N)}_{k,E}(x)+\Theta(x-L/2)\tilde{\bm \varphi}^{(\rm S-right)}_{k,E}(x)~,
\end{equation}
where $\Theta(x)$ is the Heaviside step function and we impose the continuity of the wavefuction 
at the interfaces $x=\pm L/2$
\begin{eqnarray}\label{eqn:continM}
\tilde{\bm \varphi}^{\rm (S-left)}_{k,E}(x=-L/2)&=&\tilde{\bm \varphi}^{\rm (N)}_{k,E}(x=-L/2)~,\\
\label{eqn:continP}
\tilde{\bm \varphi}^{\rm (S-right)}_{k,E}(x=L/2)&=&\tilde{\bm \varphi}^{\rm (N)}_{k,E}(x=L/2)~.
\end{eqnarray}
We start from the superconducting left-side (S-left), $x<-L/2$ and we look for a solution $\tilde{\bm \varphi}^{\rm S-left}_{k,E}(x)$ expressed as linear combination of terms the form $e^{s \eta_\lambda (x+L/2)} {\bm w}_{k,E,\lambda,s}$, where $s=\pm$, $\lambda=\pm$, and
\begin{equation}\label{eqn:seta}
 \eta_{\lambda} = i \sqrt{\left(\frac{U_0+\lambda \sqrt{E^2-\Delta_0^2}}{\hbar \vD}\right)^2-k^2}~,
\end{equation}
\begin{equation}
{\bm w}_{k,E,\lambda,s} = {\cal W}_{\lambda,s} {\bm b}_{k,E,\lambda}~,
\end{equation}
\begin{equation}
{\bm b}_{k,E,\lambda}= \frac{1}{\sqrt{2}} \left[e^{i \phi_{\rm L}} e^{i\frac{\lambda}{2}{\rm arccos}(E/\Delta_0)} ,0,e^{-i\frac{\lambda}{2}{\rm arccos}(E/\Delta_0)},0\right]
 ^{\rm T}~,
\end{equation}
\begin{equation}
{\cal W}_{\lambda,s} =Q_{\lambda,s} \Lambda_{\lambda,s}~,
\end{equation}
\begin{equation}
Q_{\lambda,s}=\frac{1}{\sqrt{2}}[\sigma_z+i(e^{-s z_\lambda} \sigma_- - e^{s z_\lambda} \sigma_+)]=Q_{\lambda,s}^{-1}~,
\end{equation}
\begin{equation}
 \Lambda_{\lambda,s} = \frac{1}{\sqrt{2[1+e^{s(z_\lambda+z^\ast_\lambda)}]}}(1-\sigma_z)
 +\frac{1}{\sqrt{2[1+e^{-s(z_\lambda+z^\ast_\lambda)}]}}(1+\sigma_z)
 ~,
\end{equation}
and
\begin{equation}
e^{s z_\lambda}=
 \frac{\hbar \vD(k+ s\eta_{\lambda})}{U_0+\lambda \sqrt{E^2-\Delta_0^2}}~.
\end{equation}
For the right-superconducting side (S-right), $x>L/2$, we take the same form with the replacement  $-L/2 \to L/2$ and  $\phi_{\rm L}\to \phi_{\rm R}$.
Assuming that the superconductive sides are in large doping regime, $U_0 \gg |\mu_0|,\Delta_0$, the complex wavenumber in Eq.~\eqref{eqn:seta} can be approximated as
\begin{equation}\label{eqn:seta2}
 \eta_{\lambda} \approx i \frac{U_0}{\hbar \vD}-\lambda\frac{ \sqrt{\Delta_0^2-E^2}}{\hbar \vD}~,
\end{equation}
where $\hbar \vD/\sqrt{\Delta_0^2-E^2}$ is the penetration length, and $z_\lambda \approx i \pi/2$.
The physical solutions in the 
left (right) superconductive side 
correspond to $s\lambda=-$ ($s\lambda=+$). 
The normalized eigenfunctions of the Andreev bound state labeled by energy $E$ and $y$-component of the wavevector $k$, 
in S-left  side  take the form 
%
\begin{eqnarray}\label{eqn:SL}
\tilde{\bm \varphi}^{\rm (S-left)}_{k,E}(x)&=& \sqrt{\frac{\sqrt{\Delta_0^2-E^2}}{\hbar \vD}}e^{ 
\sqrt{\Delta_0^2-E^2}(x+L/2)/\hbar \vD} \nonumber\\
&\times&[e^{-i U_0 (x+L/2)/(\hbar \vD)} x_{\rm L}   {\bm w}_{k,E,+,-} +e^{i U_0 (x+L/2)/(\hbar \vD)}y_{\rm L}  {\bm w}_{k,E,-,+}]~,
\end{eqnarray}
similarly, in the S-right side one has
\begin{eqnarray}\label{eqn:SR}
\tilde{\bm \varphi}^{\rm (S-right)}_{k,E}(x)&=& \sqrt{\frac{\sqrt{\Delta_0^2-E^2}}{\hbar \vD}}e^{ 
-\sqrt{\Delta_0^2-E^2}(x-L/2)/\hbar \vD} \nonumber\\
&\times&[e^{-i U_0 (x-L/2)/(\hbar \vD)} x_{\rm R}   {\bm w}_{k,E,+,+} +e^{i U_0 (x-L/2)/(\hbar \vD)}y_{\rm R}  {\bm w}_{k,E,-,-}]~,
\end{eqnarray}
%
where  $\{ x_{\rm L} ,y_{\rm L} ,x_{\rm R} ,y_{\rm R} \}$ is a set of c-numbers.

The stationary Dirac equation for ${\bm \varphi}^{\rm (N)}_{k,E}(x,y)=e^{iky}\tilde{\bm \varphi}^{\rm (N)}_{k,E}(x)$ in the
normal phase region, $|x|<L/2$, it
can be expressed in terms of a transfer matrix ${\cal T}(k,E; x)$ as
\begin{equation}
\tilde{\bm \varphi}^{\rm (N)}_{k,E}(x)={\cal T}(k,E; x)\tilde{\bm \varphi}^{\rm (N)}_{k,E}(-L/2)~, 
\end{equation}
obeying the equation
\begin{equation}\label{eq:Tmatrix_2}
\frac{d  {\cal T}(k,E; x)}{d x}=
\left [
i \tau_z \sigma_x \frac{E}{\hbar \vD} + i \sigma_x  \frac{\mu_0}{\hbar \vD} +  \sigma_z k    \right ] {\cal T}(k,E;x)~,
\end{equation}
with boundary condition
\begin{equation}
  {\cal T}(k,E; x=-L/2)= \openone_{\tau} \openone_{\sigma}~.
\end{equation}
Explicitly, the transfer matrix takes the following form
\begin{equation}
  {\cal T}(k,E;x) = \frac{1+\tau_z}{2} {\mathbb T}(k,E;x) +  \frac{1-\tau_z}{2} {\mathbb T}(k,-E;x)~,
\end{equation}
where
\begin{equation}
{\mathbb T}(k,\pm E;x)= \sin[q_\pm (x+L/2)] \left[i\frac{\mu_0\pm E}{\hbar \vD q_\pm} \sigma_x+\frac{k}{q_\pm} \sigma_z \right] + \cos[q_\pm (x+L/2)]~,
\end{equation}
and
\begin{equation}
q_\pm =\sqrt{\left(\frac{\mu_0\pm E}{\hbar \vD } \right)^2 -k^2}~. 
\end{equation}
The transfer matrix fulfills the following useful property
\begin{equation}\label{eqn:TsxT}
{\cal T}^\dagger(k,E;x) \openone_\tau \sigma_x {\cal T}(k,E;x)  = \openone_\tau \sigma_x~.
\end{equation}
The continuity conditions of the wavefunction, 
 \eqref{eqn:continM}-\eqref{eqn:continP}, can be written in compact form  as follows
\begin{equation}\label{eqn:dxTsx}
\tilde{\bm \varphi}^{\rm (S-right)}_{k,E}(x=L/2)={\cal T}(k,E;L/2)\tilde{\bm \varphi}^{\rm (S-left)}_{k,E}(x=-L/2)~.
\end{equation}
After easy algebraic manipulations, these constraints reduce to the following homogeneous system of four equations for 
$\{ x_{\rm L} ,y_{\rm L} ,x_{\rm R} ,y_{\rm R} \}$,
\begin{equation}\label{eqn:D=0}
x_{\rm R}   {\bm w}_{k,E,+,+} +y_{\rm R}   {\bm w}_{k,E,-,-}={\cal T}(k,E;L/2)[x_{\rm L}   {\bm w}_{k,E,+,-} +y_{\rm L}   {\bm w}_{k,E,-,+}]~.
\end{equation}
The solution of this system is not trivial only if the associated matrix is singular, i.e. its determinant $\cal D$ is zero. 
For each couple of  wavenumber $k$ and phase difference $\phi=\phi_{\rm R}-\phi_{\rm L}$, the energies which nullifies the determinant $\cal D$ represent the eigenenergies of the Andreev bound states.
In this work, we focus on short GJJs, $\Delta_0 \ll \hbar \vD/L$. In this limit the Andreev bound eigenenergies satisfy the following inequality $|E| \lesssim \Delta_0 \ll \hbar \vD/L$. Thus, the transfer matrix can be approximated as ${\cal T}(k,E;L/2)\approx {\cal T}(k,0;L/2)$. 
For a given pair of values $k$ and $\phi$, the determinant $\cal D$ vanishes for $j  \epsilon(k,\phi)$, where $j=\pm$ is a subband index and
\begin{eqnarray}
 \epsilon(k,\phi)&=&\Delta_0 \sqrt{1-\tau(k) \sin^2(\phi/2)}~,\\
 \tau(k)&=&\frac{\kF^2-k^2}{\kF^2-k^2 \cos ^2 (L \sqrt{\kF^2-k^2} )}~,
\end{eqnarray}
and $\kF=\mu_0/(\hbar \vD)$.
The normalized eigenfunction with subband index $j$ has components
\begin{eqnarray}
x_{L,j} &=&  - j M_{1}(k) \sin[\phi/2+j\arccos(  \epsilon(k,\phi) /\Delta_0)]C_j
\\
y_{L,j}&=&   M_{2}(k) \sin(\phi/2) C_j\\
x_{R,j}&=& -e^{-i \phi/2}\frac{\sin[\arccos(  \epsilon(k,\phi) /\Delta_0)]}{\sin(\phi/2)} M_{1}(k) y_L~,\\
y_{R,j}&=& e^{-i \phi/2}\frac{\sin[\arccos(  \epsilon(k,\phi)  /\Delta_0)]}{\sin(\phi/2)} M^\ast_{1}(k) x_L~,\\
C_j&=&1/\sqrt{2 |M_{1}(k)|^2\sin[\phi/2+j\arccos( \epsilon(k,\phi)/\Delta_0)]\sin(\phi/2)  \epsilon(k,\phi)/\Delta_0}~.
\end{eqnarray}
where
\begin{equation}
 M_{1}(k) =\cos [ \sqrt{(k_{\rm F}^2-k^2 )  } L ] + i
 \frac{ k_{\rm F}  \sin [\sqrt{(k_{\rm F}^2-k^2 )  }L]}{\sqrt{k_{\rm F}^2-k^2   } }~,
\end{equation}
and
\begin{equation}
 M_{2}(k) =
 \frac{ k   \sin [\sqrt{(k_{\rm F}^2-k^2 )  }L]}{\sqrt{k_{\rm F}^2-k^2   } }~.
\end{equation}

In the central normal phase region, $|x|<L/2$, in the short junction limit the wavefunction of the Andreev bound state is expressed in the following spinorial form
\begin{equation}\label{eqn:TN}
{\bm \varphi}^{\rm (N)}_{k,j \epsilon(k,\phi)}(x) = {\cal T}(k,0;x)[x_{{\rm L},j}   {\bm w}_{k,j \epsilon(k,\phi),+,-} +y_{{\rm L},j}   {\bm w}_{k,j \epsilon(k,\phi),-,+}]~.
\end{equation}
Within the Dirac-Bogoliubov-De Gennes formalism the charge density operator and the current density operator are  expressed as
\begin{eqnarray}
\hat{\rho}_{\rm e} &=& -e  \sum_{\zeta=\pm} \hat{\Psi}^\dagger_{\zeta}(\br) \tau_z \openone_\sigma  \hat{\Psi}_{\zeta}(\br)~, \\
\hat{\bm J}_{\rm e}(\br) &=& -e \vD  \sum_{\zeta=\pm} \hat{\Psi}^\dagger_{\zeta}(\br) \openone_\tau  {\bm \sigma}  \hat{\Psi}_{\zeta}(\br)~,
\end{eqnarray}
they are both diagonal in the valley index $\zeta=\pm$.
The charge  current operator which describes the charge flow through the normal phase region along the $x$-direction reads
\begin{equation}
\hat{ I}(x) = \int^{W/2}_{-W/2} d y {\bm u}\cdot \hat{\bm J}_{\rm e}(\br)~,
\end{equation}
where ${\bm u}=-{\bm e}_x$ is the unit vector.
Within the normal phase region, 
we evaluate the matrix element of the charge current operator as
\begin{equation}
 {I}_{(k,j),(k^\prime,j^\prime)}(x)=  
 e \vD
\int^{W/2}_{-W/2} \frac{d y}{W} e^{-i(k-k^\prime)y}{\bm \varphi}^{\rm (N)\dagger}_{k,j \epsilon(k,\phi)}(x) \openone_\tau \sigma_x  {\bm \varphi}^{\rm (N)}_{k^\prime,j^\prime \epsilon(k^\prime)}(x)~,
\end{equation}
for large $W$
\begin{equation}
 { I}_{(k,j),(k^\prime,j^\prime)}(x)=
 e \vD \delta_{k,k^\prime}
 {\bm \varphi}^{\rm (N)\dagger}_{k,j \epsilon(k,\phi)}(x) \openone_\tau \sigma_x  {\bm \varphi}^{\rm (N)}_{k,j^\prime  \epsilon(k,\phi)}(x)~.
\end{equation}
By using the property of the transfer matrix in Eq.\eqref{eqn:TsxT} for the spinor in Eq.\eqref{eqn:TN}, we obtain that the matrix element ${ I}_{(k,j),(k^\prime,j^\prime)}(x)$ has the same value for each $x$ position inside the normal phase region.
By focusing on the Andreev bound states, the current operator can be written in the following compact form 
\begin{equation}\label{eqn:SIA}
\hat{ I}_{\rm A}=
-
\frac{e \Delta_0^2}{\hbar} \sum_{\zeta=\pm} \sum_k \frac{\tau(k)}{\ \epsilon(k,\phi)} \sin(\phi/2)  [\cos(\phi/2) \hat{\Sigma}^z_{\zeta,k} - \sqrt{1-\tau(k)} \sin(\phi/2) \hat{\Sigma}^x_{\zeta,k}]~,
\end{equation}
where $\hat{\Sigma}^z_{\zeta,k}=\hat{\gamma}_{+,k,\zeta}^\dagger \hat{\gamma}_{+,k,\zeta}-\hat{\gamma}_{-,k,\zeta}^\dagger \hat{\gamma}_{-,k,\zeta}$,
$\hat{\Sigma}^x_{\zeta,k}=\hat{\gamma}_{+,k,\zeta}^\dagger \hat{\gamma}_{-,k,\zeta}+\hat{\gamma}_{-,k,\zeta}^\dagger \hat{\gamma}_{+,k,\zeta}$, 
$\hat{\gamma}_{+,k,\zeta}$ and $\hat{\gamma}_{-,k,\zeta}$ are respectively the upper and lower Andreev bound state fermionic operator labeled by the $y$-component of the wavevector $k$ and the valley index $\zeta$.

\section{Supplementary Note 2}

We analyze the tunneling Hamiltonian which describes the interaction between a single impurity and the electron gas in a graphene monolayer. 
We describe the electron gas in graphene in the tight binding model~\cite{katsnelson_book_2012}. The impurity consists of a single energy level, and the tunneling Hamiltonian is expressed as
\begin{equation}
\hat{V}_{t}= \sum_{\bm j} \sum_{\alpha=A,B} t_{\alpha,{\bm j}} \ket{d}\bra{\bm R_{\bm j},\alpha}+{\rm h.c.}~,
\end{equation}
\begin{equation}
t_{\alpha,{\bm j}}=
\int d^2 \br \psi_d^\ast(\br) {\cal V}_{ t}(\bm{r})  \phi_{\alpha}(\bm{r}-\bm{R}_{\bm j}) ~,
\end{equation}
where ${\bm j}=(j_1,j_2)^{\rm T}$ is a vector composed by two integer components,
$t_{\alpha,{\bm j}}$ is assumed to be real, since both wavefuctions are taken real without loss of generality.
Indeed, the atomic wavefunction $\phi_{\alpha}(\bm{r}-\bm{R}_{\bm j})$ describes the {\it p}$_z$ orbitals of the carbon atoms centered in site $\bm R_{\bm j}$ of the sublattice $\alpha$ ($\alpha=A,B$)~\cite{bena_newjp_2009}, and the bound states can be
described by real wavefunctions, $\psi_d^\ast(\br)$~\cite{griffiths_book_2018}.
The lattice vectors are expressed in terms of
the unit vectors as ${\bm R}_{\bm j}=j_1 {\bm a}_1+ j_2 {\bm a}_2$, where
${\bm a}_1 = a[3/2,\sqrt{3}/2]^{\rm T}$, ${\bm a}_2 = a[3/2,-\sqrt{3}/2]^{\rm T}$, and $a=1.42$~\AA~is the carbon-carbon distance.
Here, we focus on the low-energy physics of the electron system in graphene. We expand the tunneling Hamiltonian close to the Dirac points $\bm K=[2\pi/(3 a),2\pi\sqrt{3}/(3 a)]^{\rm T}$ and $\bm K^\prime=-\bm K$,
\begin{equation}
\hat{V}=\sum_{\zeta=\pm} \sum_{\alpha=A,B} \sum^\prime_\bq 
\tilde{t}^\zeta_{\alpha}(\bm{q}) \ket{d}\bra{\zeta \bm{K}+\bq,\alpha}+
\tilde{t}^{\zeta \ast}_{\alpha}(\bm{q})  \ket{\zeta \bm{K}+\bq,\alpha}\bra{d}~,
\end{equation}
the sum with the prime symbol is limited to $|\bm{q}|\ll \pi/a$.
The tunneling matrix element is expressed as
\begin{equation}
\tilde{t}^\zeta_{\alpha}(\bm{q}) = \sum_{\bm j} \frac{e^{i \zeta \bm{K}\cdot \bm{R}_{\bm j}}}{\sqrt{N}}e^{i \bq \cdot \bm{R}_{\bm j}}
t_{\alpha,{\bm j}}~,
\end{equation}
where $N$ is the number of the unit cells in the graphene sample.  In the long-wavelength approximation, the tunneling Hamiltonian reads
\begin{equation}
\hat{V}=\sum_{\zeta=\pm} \sum_{\alpha=A,B} \int d \br
t^\zeta_{\alpha}(\bm{r}) \ket{d}\bra{\zeta,\alpha,\br}+
t^{\zeta \ast}_{\alpha}(\bm{r})  \ket{\zeta,\alpha,\br}\bra{d}~,
\end{equation}
where
\begin{equation}
t^\zeta_{\alpha}(\bm{r})=\frac{1}{\sqrt{N A_{\rm c}}} \sum^\prime_{\bq} 
e^{- i \bq \cdot \br}
\tilde{t}^\zeta_{\alpha}(\bm{q})~, 
\end{equation}
\begin{equation}
\ket{\zeta,\alpha,\br}=\frac{1}{\sqrt{N A_{\rm c}}} \sum^\prime_{\bq} 
e^{- i \bq \cdot \br}
\ket{\zeta \bm{K}+\bm{q},\alpha}~,
\end{equation}
$A_{\rm c}=3\sqrt{3}a^2/2$ is the area of a unit cell~\cite{katsnelson_book_2012}.
If the tunneling term has the short-range form
\begin{equation}
 t_{\alpha,{\bm j}}=t_0 \delta_{\alpha,A} \delta_{\bm j,\bm j_0}~,
\end{equation}
the corresponding tunneling matrix element in the long-wavelength approximation reads
\begin{equation}
 t^\zeta_{\alpha}(\bm{r})=t_0 \delta_{\alpha,A} 
 \frac{1}{\sqrt{N A_{\rm c}}} \sum^\prime_{\bq} 
e^{- i \bq \cdot \br}
\frac{e^{i \zeta \bm{K}\cdot \bm{R}_{j_0}}}{\sqrt{N}}e^{i \bq \cdot \bm{R}_{\bm j_0}}~.
\end{equation}
By using the approximation
\begin{equation}
\sum^\prime_{\bm q} e^{i \bq\cdot(\br-\br^\prime)}
\approx N A_{\rm c} \delta(\br-\br^\prime)~,
\end{equation}
one has
\begin{equation}
 t^\zeta_{\alpha}(\bm{r})\approx t_0 \sqrt{A_{\rm c}} \delta_{\alpha,A} 
e^{i \zeta \bm{K}\cdot \bm{R}_{\bm j_0}} \delta(\br - {\bm R}_{\bm j_0})
=t_0 \sqrt{A_{\rm c}} \delta_{\alpha,A} 
e^{-i \zeta (j_{0,1}-j_{0,2})\frac{2\pi}{3}} \delta(\br - {\bm R}_{\bm j_0})
~,
\end{equation}
where the dependence on sublattice $\alpha$ appears only in the Kronecker delta,
the phase factor can take three values, i.e. $e^{-i 2 \zeta n_d \pi/3}$ with $n_d=-1,0,1$.
In a region centered in $\bm{r}_d$ of area $\lambda^2$, where $\lambda \gg a$, such that $2\pi/\lambda$ is the cut-off wavenumber of the long-wavelength approximation, there is a large number of lattice sites.
In the long-wavelength approximation, in a region centered in $\bm{r}_d$ of area $\lambda^2$ one can not distinguish the exact lattice site where a impurity acts, and the tunneling matrix element is expressed as
\begin{equation}
 t^\zeta_{\alpha}(\bm{r})=t_0 \sqrt{A_{\rm c}} [m_d \delta_{\alpha,A} +(1-m_d)\delta_{\alpha,B}]
e^{-i \zeta n_d \frac{2\pi}{3}} \delta(\br - {\bm r}_{d})~,
\end{equation}
where the indices $m_d$ and $n_d$ can assume the values $\{0,1\}$ and $\{-1,0,1\}$, respectively.  

\section{Supplementary Note 3}

Given an Hamiltonian partitioned in blocks as
\begin{equation}
H= 
\begin{bmatrix}
 H_{\rm PP}& H_{\rm PQ}\\
H_{\rm QP}&  H_{\rm QQ}
\end{bmatrix}~,
\end{equation}
the Green's functions can be defined blockwise as follows
\begin{equation}
\begin{bmatrix}
 \Omega - H_{\rm PP}& -H_{\rm PQ}\\
-H_{\rm QP}&  \Omega - H_{\rm QQ}
\end{bmatrix}
\begin{bmatrix}
G_{\rm PP}&G_{\rm PQ}\\
G_{\rm QP}&G_{\rm QQ}
\end{bmatrix}
=
\begin{bmatrix}
\openone_{\rm PP}&0\\
0&\openone_{\rm QQ}
\end{bmatrix}~,
\end{equation}
where
\begin{equation}\label{eq:Gpp}
 G_{\rm  PP}=\{\Omega-H_{\rm PP} - H_{\rm PQ}[\Omega-H_{\rm QQ}]^{-1}H_{\rm QP}\}^{-1}~,
\end{equation}
\begin{equation}
 G_{\rm QP}=[ \Omega - H_{\rm QQ}]^{-1}H_{\rm QP} G_{\rm PP}~,
\end{equation}
\begin{equation}
 G_{\rm PQ}= G_{\rm PP} H_{\rm PQ}[\Omega-H_{\rm QQ}]^{-1}~,
\end{equation}
\begin{equation}
 G_{\rm QQ}=[\Omega-H_{\rm QQ}]^{-1}
+[\Omega-H_{\rm QQ}]^{-1}H_{\rm QP}  G_{\rm PP} H_{\rm PQ}[\Omega-H_{\rm QQ}]^{-1}~.
\end{equation}

\section{Supplementary Note 4}

Given a generic operator $\hat{\cal O}$ associated with a measurable quantity, the thermal average value is defined as
\begin{equation}
\langle \hat{\cal O}(t)\rangle = \frac{{\rm Tr}[ e^{-\beta \hat{\cal H}} e^{i\hat{\cal H} t/\hbar} \hat{\cal O} e^{-i\hat{\cal H} t/\hbar} ]}{{\rm Tr}[e^{-\beta \hat{\cal H}}]}~,
\end{equation}
where $\beta=(k_{\rm B} T)^{-1}$. For time-independent Hamiltonians the thermal average $\langle \hat{\cal O}(t)\rangle$ is stationary.
By resolving the thermal average in the energy eigenstates, one has
\begin{equation}
 \langle \hat{\cal O} (t)\rangle=\langle \hat{\cal O}\rangle=\sum_\lambda n_{\rm F}(E_{\lambda}) \bra{\lambda}  \hat{\cal O} \ket{\lambda}~,
\end{equation}
where $n_{\rm F}(x)=[1+\exp(\beta x)]^{-1}$ is the Fermi-Dirac function.
After easy algebraic manipulations, the thermal average  $\langle \hat{\cal O}\rangle$ can be expressed in a generic basis as
\begin{equation}\label{eq:Tr0v3}
  \braket{\hat{\cal O} }=
  -\int  \frac{d \Omega}{2 \pi}  \sum_{\nu,\nu^\prime} n_F(\Omega)  {\cal O}_{\nu \nu^\prime}  G_{R-A}(\nu^\prime,\nu,\Omega)
  ~,
\end{equation}
where
\begin{equation}\label{eq:Gra}
G_{R-A}(\nu^\prime,\nu,\Omega)= G_R(\nu^\prime,\nu,\Omega)-G_A(\nu^\prime,\nu,\omega)~,
\end{equation}
\begin{equation}
G_R(\nu^\prime,\nu,\Omega)=\braket{\nu^\prime|\frac{1}{\Omega+i0^+-\hat{\cal H}}|\nu}=\sum_\lambda  \frac{ \braket{\nu^\prime|\lambda} \braket{\lambda|\nu}}{\Omega+i0^+-E_\lambda}~,
\end{equation}
\begin{equation}
G_A(\nu^\prime,\nu,\omega)=\braket{\nu^\prime|\frac{1}{\Omega-i0^+-\hat{\cal H}}|\nu}=\sum_\lambda  \frac{ \braket{\nu^\prime|\lambda} \braket{\lambda|\nu}}{\Omega-i0^+-E_\lambda}~.
\end{equation}
If we choose a basis set $\ket{\nu}$ where the Green's functions $ G_{R/A}(\nu^\prime,\nu,\Omega)$ are diagonal, one can write
\begin{equation}
 G_{R-A}(\nu^\prime,\nu,\Omega)=\delta_{\nu^\prime,\nu} A(\nu,\Omega)~,
\end{equation}
where
\begin{equation}
 A(\nu,\Omega)=-2 {\rm Im} G_R(\nu,\nu,\Omega)~
\end{equation}
is the spectral function.

Similarly, we define the correlation function
\begin{eqnarray}
S_{{\cal O}}(t',t)&=&\langle \hat{\cal O}(t') \hat{\cal O}(t)\rangle-\langle \hat{\cal O}(t')\rangle \langle \hat{\cal O}(t)\rangle\\
&=& \frac{{\rm Tr}[ e^{-\beta {\cal H}} e^{i{\cal H} t'/\hbar} {\cal O} e^{-i{\cal H} (t'-t)/\hbar} {\cal O} e^{-i{\cal H} t/\hbar}]}{{\rm Tr}[e^{-\beta {\cal H}}]}-  \braket{{\cal O} }^2~.\nonumber
\end{eqnarray}
For a time independent Hamiltonian one has $S_{{\cal O}}(t',t)=S_{{\cal O}}(t'-t)$, and in the energy basis 
\begin{equation}
 S_{{\cal O}}(t'-t)=
 \sum_{\lambda ,\lambda^\prime} n_F(E_\lambda)[1-n_F(E_{\lambda^\prime})] {\cal O}_{\lambda,\lambda^\prime} {\cal O}_{\lambda^\prime, \lambda} e^{-i(E_{\lambda^\prime}-E_\lambda)(t'-t)/\hbar}~.
\end{equation}
The correlation function can be expressed in a generic basis as
\begin{eqnarray}\label{eq:OOv2}
 S_{{\cal O}}(t'-t)&=&
\int  \frac{d \Omega}{2 \pi} \int  \frac{d \Omega^\prime}{2 \pi} \sum_{\nu_1,\nu_2,\nu_3,\nu_4}  e^{-i(\Omega^\prime-\Omega)(t'-t)/\hbar}
 \\
&\times& n_F(\Omega) [1-n_F(\Omega^\prime)]{\cal O}_{\nu_1 \nu_2} {\cal O}_{\nu_3 \nu_4} G_{R-A}(\nu_4,\nu_1,\Omega)  G_{R-A}(\nu_2,\nu_3,\Omega^\prime)~,  \nonumber
\end{eqnarray}
leading to the following form in the frequency domain 
\begin{eqnarray}\label{eq:SOO}
S_{{\cal O}}(\omega)&=& \int^\infty_{-\infty} dt e^{i \omega(t'-t)}S_{{\cal O}}(t'-t)=
\hbar \int  \frac{d \Omega}{2 \pi}  \sum_{\nu_1,\nu_2,\nu_3,\nu_4}
 \\
&\times& n_F(\Omega) [1-n_F(\Omega+\hbar\omega)]{\cal O}_{\nu_1 \nu_2} {\cal O}_{\nu_3 \nu_4} { G}_{R-A}(\nu_4,\nu_1,\Omega)  {G}_{R-A}(\nu_2,\nu_3,\Omega+\hbar \omega)~.  \nonumber
\end{eqnarray}

\newpage


\end{document}